%% file: FinalSubmissionHuleihel_Steinberg.tex
\documentclass[conference]{IEEEtran}
\usepackage{mathrsfs}  
\usepackage{graphics}
\usepackage[dvips]{graphicx}
\usepackage[cmex10]{amsmath}
\usepackage[multiple]{footmisc}
\usepackage{amssymb}

\input{myshorts}

\usepackage{epsf}
\usepackage{epsf}
\usepackage{calc}
\usepackage{graphicx}
\usepackage{psfrag}

\usepackage{enumitem}
\usepackage{lipsum}
\usepackage[usenames,dvipsnames]{pstricks}
 \usepackage{epsfig}
 \usepackage{pst-grad} 
 \usepackage{pst-plot} 
\newcounter{MYtempeqncnt}
\setitemize[itemize]{leftmargin=*,itemindent=10pt}
\begin{document}

\def\naive{na\"{\i}ve}
\newtheorem{problem}{Problem}
\newtheorem{definition}{Definition}
\newtheorem{lemma}{Lemma}
\newtheorem{proposition}{Proposition}
\newtheorem{corollary}{Corollary}
\newtheorem{example}{Example}
\newtheorem{conjecture}{Conjecture}
\newtheorem{algorithm}{Algorithm}
\newtheorem{theorem}{Theorem}
\newtheorem{exercise}{Exercise}
\newtheorem{remark}{Remark}

\newcommand{\bml}[1]{\mbox{\boldmath $ #1 $}}
\newcommand{\ssbml}[1]{\scriptsize{\mbox{\boldmath $ #1 $}}}
\newcommand{\sbml}[1]{\mbox{\scriptsize{\boldmath $ #1 $}}}
\newcommand{\p}[1]{\left(#1\right)}
\newcommand{\pp}[1]{\left[#1\right]}
\newcommand{\ppp}[1]{\left\{#1\right\}}
\newcommand{\norm}[1]{\left\|#1\right\|}
\newcommand{\abs}[1]{\left|#1\right|}
\newcommand{\limn}{\lim_{n\rightarrow\infty}}

\newcommand{\one}{\frac{1}{n}}
\newcommand{\half}{\frac{1}{2}}
\newcommand{\onei}{{\rm 1\!\!\!\:I}}
\newcommand{\NN}{{\rm I\!\!\!\;N}}
\newcommand{\EE}{{\rm I\!\!\!\;E}}

\renewcommand{\IEEEQED}{\IEEEQEDopen}

\newcommand{\markov}{\mbox{$-\hspace{-2.3mm}\circ\hspace{1mm}$}}
\def\squarebox#1{\hbox to #1{\hfill\vbox to #1{\vfill}}}
\newcommand{\qed}{\hspace*{\fill}
            \vbox{\hrule\hbox{\vrule\squarebox{.667em}\vrule}\hrule}\smallskip}


\title{Channels with Cooperation Links that May Be Absent}

\author{
\IEEEauthorblockN{Wasim Huleihel\IEEEauthorrefmark{1} and Yossef Steinberg\IEEEauthorrefmark{2}}\\
\IEEEauthorblockA{\IEEEauthorrefmark{1}Research Laboratory of Electronics\\
Massachusetts Institute of Technology\\
Cambridge, MA\\
{\tt wasimh@mit.edu} 
}
\IEEEauthorblockA{\IEEEauthorrefmark{2}Andrew \& Erna Viterbi Faculty of Electrical Engineering\\
Technion - Israel Institute of Technology\\
Haifa 32000, Israel\\
{\tt ysteinbe@ee.technion.ac.il} 
}
}

\maketitle

\begin{abstract}
  It is well known  that cooperation between users in a communication network
  can lead to significant performance gains. A common assumption in past works is that
  all the users are aware of the resources available for cooperation, and know exactly 
  to what extent these resources can be used. Unfortunately, in many modern communication networks
  the availability of cooperation links cannot be guaranteed a priori, due to the dynamic
  nature of the network. In this work a family 
  of models is suggested where the cooperation links may or may not be present.
  Coding schemes are devised that exploit the cooperation links if they are present, 
  and can still operate (although at reduced rates) if cooperation is not possible.
\end{abstract}

\section{Introduction}
\label{sec:intro}
Communication\makeatletter{\renewcommand*{\@makefnmark}{}
\footnotetext{This paper was presented in part at the 2014 and 2016 IEEE International Symposium on Information Theory. This work was supported by the  ISRAEL SCIENCE FOUNDATION (ISF) (grant no. 684/11).}\makeatother}
techniques that employ cooperation between users in a network have been an extensive area of research 
in recent years. The interest in such schemes stems from the potential increase in the network performance. 
The employment of cooperative schemes require the use of system resources - 
bandwidth, time slots, energy, etc - that should be allocated for the cooperation to take place. 
Due to the dynamic nature of modern, wireless ad-hoc communication systems, the availability
of these resources is not guaranteed a priori, as they depend on parameters that the system designer does not have any control on. 
For example, the cooperation can depend on the battery status of intermediate users (relays), 
on weather, or just on the willingness of peers in the network to help. 
A typical situation, therefore, is that the users are aware of the \emph{possibility} that cooperation will take place, 
but it cannot be assured before transmission begins. Moreover, in many instances it is not possible 
to inform the transmitter whether or not a potential relay/helper decides to help. 
Thus, the traditional approach leaves  the designer with two design options. 
Option 1 is the pessimistic one: 
assume that none of the unreliable relays exists, and design a system without cooperation. 
Option 2 takes the optimistic view:
assume that the potential relays exist, and design a system with cooperation. 
The pros and cons of each of the designs are clear. 
Option 1 is ``safe,'' but results in relatively low rates. 
Option 2 aims to transmit at the maximal rates, but runs the risk that some of the relays/helpers are absent, 
in which case the coding scheme collapses.

In this work we suggest a third option: design a robust system,
which takes advantage of the links if they are present, but can operate also if they are absent, 
although possibly at reduced rates. 
This design problem becomes simple if all the users in the system can be informed  
about the situation of the helpers before transmission begins. 
We study models in which, at least for part of the users, 
this information is not available before transmission. In general, this set of problems can be viewed as 
channel coding counterparts of well known problems in source coding, like multiple descriptions \cite{GamalCover,FuYeung}, 
successive refinement \cite{EquitzCover,Rimoldi}, and rate-distortion when side information may be absent \cite{J_HeegardBerger85Rate}.
We focus on two channel models - the physically degraded broadcast channel (BC) with conferencing decoders, and the
multiple access channel (MAC) with cribbing encoders. The BC with conferencing decoders
was first studied by Dabora and Servetto~\cite{C_DaboraServetto04Broadcast}, \cite{J_DaboraServetto06Broadcast},
and independently by Liang and Veeravalli~\cite{C_LiangVeeravalli04Impact}, \cite{J_LiangVeeravalli07Cooperative},
who studied also the more general setting of relay-broadcast channels (RBC). 
In  the model of Dabora and Servetto, a two-users BC is considered, where
the decoders can exchange information 
via noiseless communication links of limited capacities $C_{1,2}$ and $C_{2,1}$. 
When the broadcast channel is physically degraded, information sent from the weaker (degraded) user to the stronger
is redundant, and only the capacity of the link from the stronger user to the weaker (say $C_{1,2}$)
increases the communication rates. For this case, Dabora and Servetto characterized the capacity region. 
Their result coincides with the results of Liang and Veeravalli when the relay link
of~\cite{C_LiangVeeravalli04Impact} is replaced with a constant rate bit pipe.

The MAC with cribbing encoders was introduced by Willems and Van~Der~Meulen in~\cite{J_WillemsMeulen85Discrete}.
Here there is no dedicated communication link that can be used explicitly for cooperation. Instead,
one of the encoders can crib, or listen, to the channel input of the other user. This model describes
a situation in which users in a cellular system are located physically close to each other, 
enabling part of them to listen to the transmission of the others
with high reliability - i.e., the channel  between  the transmitters that are located in close 
vicinity is almost noiseless. Willems and Van~Der~Meulen considered in~\cite{J_WillemsMeulen85Discrete} 
all consistent scenarios of cribbing (strictly causal, causal, non-causal, and symmetric or asymmetric),
and characterized the capacity region of these models. 
Another relevant recent work is \cite{PermuterAsnani}, 
where the MAC with partial cribbing encoders was considered, 
motivated by the additive noise Gaussian MAC model, where
perfect cribbing means full cooperation between the encoders and requires an infinite entropy link. 
Finally, we mention \cite{OzgurKolte}, which considers the MAC channel with state and cribbing encoders. 
Accordingly, the state can be specialized to capture the availability of the cribbing links, 
which would lead to a setup similar to the one considered in our paper. 
Nonetheless, different from our paper, this state information is assumed to be causally known 
at the cribbed encoder and not to the cribbing encoder.

In the next sections, we propose and study extensions of the two models described above, 
when the cooperation links ($C_{1,2}$ of the physically degraded BC, and the cribbing link of the MAC) 
may or may not be present. For the MAC models, 
we first propose achievable rate regions which are based on the combination of 
superposition coding and block-Markov coding. 
Here, we consider the unreliable strictly causal, causal, and non-causal cribbing. 
Then, we propose a general outer bound, which is tight for some interesting special case 
where a constraint on the rates of the users is added. For the physically degraded BC, 
the results are conclusive. 
The results derived here were partially presented in \cite{SteinbergISIT2014,WasimYossi}.

It should be noted that multi-user communication systems with uncertainty in part of the network links have been
 studied in the literature - 
 see, e.g.,~\cite{J_SimeoneLevySanderovichSomekhZaidelPoorShamai12Cooperative}
 and~\cite{J_KarasikSimeoneShamai13Robust}, and references therein. 
 The models suggested here, of the BC and MAC
with uncertainty in the cooperation links, have not been studied before. 

The outline of the rest of the paper is as follows. In Section~\ref{sec:notation}, we establish our notation. 
The physically degraded BC with unreliably cooperating decoders is presented and sutdied
in Section~\ref{sec:PDBC_cooperating}. In Section~\ref{sec:MAC_cribbing}, 
we consider the MAC with cribbing encoders where the cribbing link may be absent.
The proofs are provided in Section~\ref{sec:proofs}.

\section{Notation Conventions}\label{sec:notation}
We use $H(\cdot)$ to denote the entropy of a discrete random variable (RV), and $I(\cdot;\cdot)$ to denote the mutual information between two discrete RVs. Calligraphic letters denote (discrete and finite) sets, e.g., $\calX$, the complement of $\calX$ is denoted by $\calX^c$, while $\abs{\calX}$ stands for its cardinality. The $n$-fold Cartesian product of $\calX$ is denoted by $\calX^n$. An element of $\calX^n$ is denoted by $x^n=(x_1,x_2,\ldots,x_n)$; whenever the dimension $n$ is clear from the context, vectors (or sequences) are denoted by boldface letters, e.g., $\bx$. We denote RVs with capital letters-$X$, etc. We denote by $T_\epsilon^{n}(X)$ the weakly typical set for the (possibly vector) RV $X$, see \cite{B_CsiszarKorner81Information} for the definition of this set. Finally, we denote the probability distribution of the RV $X$ over $\calX$ with $P_X$ and the conditional distribution of $Y$ given $X$ with $P_{Y|X}$.
\section{The Physically Degraded Broadcast Channel with Cooperating Decoders}
\label{sec:PDBC_cooperating}
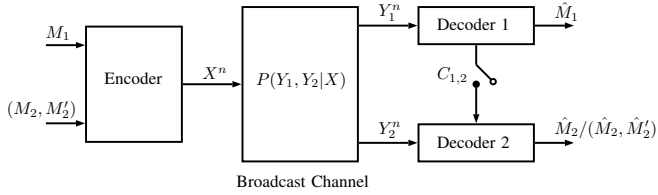
\begin{figure}[!t]
\centering
\psscalebox{0.65 0.65}
{
\begin{pspicture}(0,-1.8)(13.3,1.8)
\psframe[linecolor=black, linewidth=0.04, dimen=outer](3.6,1.4)(1.6,-1.0)
\psframe[linecolor=black, linewidth=0.04, dimen=outer](7.2,1.8)(4.8,-1.4)
\psframe[linecolor=black, linewidth=0.04, dimen=outer](10.8,1.8)(8.4,1.0)
\psframe[linecolor=black, linewidth=0.04, dimen=outer](10.8,-0.6)(8.4,-1.4)
\psline[linecolor=black, linewidth=0.04, arrowsize=0.05291666666666667cm 2.0,arrowlength=1.4,arrowinset=0.0]{->}(3.6,0.2)(4.8,0.2)
\psline[linecolor=black, linewidth=0.04, arrowsize=0.05291666666666667cm 2.0,arrowlength=1.4,arrowinset=0.0]{->}(7.2,-1.0)(8.4,-1.0)
\psline[linecolor=black, linewidth=0.04, arrowsize=0.05291666666666667cm 2.0,arrowlength=1.4,arrowinset=0.0]{->}(7.2,1.4)(8.4,1.4)
\psline[linecolor=black, linewidth=0.04, arrowsize=0.05291666666666667cm 2.0,arrowlength=1.4,arrowinset=0.0]{->}(10.8,1.4)(11.6,1.4)
\psline[linecolor=black, linewidth=0.04, arrowsize=0.05291666666666667cm 2.0,arrowlength=1.4,arrowinset=0.0]{->}(10.8,-1.0)(11.6,-1.0)
\psline[linecolor=black, linewidth=0.04, dotsize=0.07055555555555555cm 2.0,arrowsize=0.05291666666666667cm 2.0,arrowlength=1.4,arrowinset=0.0]{*->}(9.6,0.2)(9.6,-0.6)
\psline[linecolor=black, linewidth=0.04, arrowsize=0.05291666666666667cm 2.0,arrowlength=1.4,arrowinset=0.0]{->}(0.8,1.0)(1.6,1.0)
\psline[linecolor=black, linewidth=0.04, arrowsize=0.05291666666666667cm 2.0,arrowlength=1.4,arrowinset=0.0]{->}(0.8,-0.6)(1.6,-0.6)
\rput[bl](0.8,1.1){$M_1$}
\rput[bl](0.0,-0.5){$(M_2,M_2')$}
\rput[bl](2.0,0.2){Encoder}
\rput[bl](4.0,0.3){$X^n$}
\rput[bl](5.1,0.1){$P(Y_1,Y_2\vert X)$}
\rput[bl](7.6,1.5){$Y_1^n$}
\rput[bl](7.6,-0.9){$Y_2^n$}
\rput[bl](8.8,1.3){Decoder 1}
\rput[bl](11.2,1.5){$\hat{M}_1$}
\rput[bl](8.8,-1.1){Decoder 2}
\rput[bl](11.2,-0.9){$\hat{M}_2/(\hat{M}_2,\hat{M}_2')$}
\rput[bl](8.8,0.2){$C_{1,2}$}
\psline[linecolor=black, linewidth=0.04](9.6,1.0)(9.6,0.6)(9.6,0.6)
\psline[linecolor=black, linewidth=0.04, dotsize=0.07055555555555555cm 2.0]{-oo}(9.6,0.6)(10.0,0.2)
\rput[bl](4.7,-1.9){Broadcast Channel}
\end{pspicture}
}
\caption{Broadcast channel with unreliable cooperating decoders.}
\label{fig:1}

\end{figure}

Let ${\cal X}$, ${\cal Y}_1$, ${\cal Y}_2$ be finite sets. 
A broadcast channel (BC) $({\cal X},{\cal Y}_1,{\cal Y}_2,P_{Y_1,Y_2|X})$ is a channel
with input alphabet ${\cal X}$, two output alphabets ${\cal Y}_1$ and ${\cal Y}_2$, and a transition probability
$P_{Y_1,Y_2|X}$ from ${\cal X}$ to ${\cal Y}_1\times{\cal Y}_2$. The BC is said to be physically degraded if for any input distribution $P_X$, the Markov chain 
$X\markov Y_1 \markov Y_2$ holds, i.e.,
\begin{equation}
P_{X,Y_1,Y_2}=P_XP_{Y_1,Y_2|X}=P_X P_{Y_1|X} P_{Y_2|Y_1}. \label{eq:deg}
\end{equation}
We will refer to $Y_1$ (resp. $Y_2$) as the stronger (resp. weaker, or degraded) user.
We assume throughout that the channel is memoryless and that no feedback
is present, implying that the transition probability
of $n$-sequences is given by
\begin{equation}
P_{Y_1,Y_2|X}(y_1^n,y_2^n|x^n) = \prod_{i=1}^n 
P_{Y_1,Y_2|X}(y_{1,i},y_{2,i}|x_i).
\label{eq:memorylessnofeedback}
\end{equation}
Fix the transmission length, $n$, and an integer $\nu_{1,2}$. 
Let ${\cal N}_{1,2}=\left\{1,2,\ldots,\nu_{1,2}\right\}$ be the index set of the conference message.
Denote the sets of messages by ${\cal N}_k=\left\{1,2,\ldots,\nu_k\right\}$, $k=1,2$, and
${\cal N}'_2=\left\{1,2,\ldots,\nu'_2\right\}$
where $\nu_1$, $\nu_2$ and $\nu'_2$ are integers. A code for the BC
with unreliable conference link, that may or may not be present, operates as follows. 
Three messages $M_1$, $M_2$, and $M'_2$ are drawn uniformly and independently
from the sets ${\cal N}_1$, ${\cal N}_2$, and ${\cal N}'_2$, respectively. The encoder
maps this triplet to a channel input sequence, ${\bml x}(M_1,M_2,M'_2)$. 
At the channel output, Decoder $k$ has the output sequence $Y^n_k$,
$k=1,2$, at hand. Decoder 1 (resp. Decoder 2) is required to decode
the message $M_1$ (resp. $M_2$), whether or not the conference
link is present. If the conference link is present,
Decoder 1 sends a message $c\in{\cal N}_{1,2}$ to Decoder 2,
 based on the output sequence $Y_1^n$.
I.e., $c=c(Y_1^n)$. Finally, Decoder 2 decodes $M_2'$ based on his output $Y_2^n$
and the conference message $c(Y_1^n)$. The setting of the problem is depicted in Fig. \ref{fig:1}.

Observe that only Decoder 2 benefits when the conference link is present. 
Indeed, since there is only a link from Decoder 1 to Decoder 2, whatever Decoder 1 can do
with the link, he can also do without it. Therefore the rate to User 1 is independent
of whether the link is present or not. Only User 2 can benefit from its existence,
and thus there are two sets of messages intended to User 2 - 
${\cal N}_2$ and ${\cal N}_2'$. 

In the following, we give a more formal description of the above described structure. 
\begin{definition}
\label{def:BC_code}
An $(n,\nu_1,\nu_2,\nu_2',\nu_{1,2},\epsilon)$ code for the BC $P_{Y_1,Y_2|X}$ with an unreliable conference
link is an encoder mapping
\[
f: {\cal M}_1\times{\cal M}_2\times{\cal M}'_2 \rightarrow {\cal X}^n,
\]
a conference mapping
\[
h:{\cal Y}_1^n\rightarrow {\cal N}_{1,2},
\]
and three decoding maps:
\begin{subequations}
\label{eq:decoders_def}
\begin{IEEEeqnarray}{rCl}
& &    g_1: {\cal Y}_1^n\rightarrow {\cal N}_1, \label{eq:decoders_def_1}\\
& &    g_2: {\cal Y}_2^n\rightarrow {\cal N}_2, \label{eq:decoders_def_2}\\
& &    g'_2: {\cal Y}_2^n\times{\cal N}_{1,2}\rightarrow {\cal N}'_2, \label{eq:decoders_def_prime}
\end{IEEEeqnarray}
\end{subequations}
such that the average probabilities of error $P_e$ and $P_e'$ do not exceed $\epsilon$.
Here,
\begin{subequations}
\label{eq:prob_error_def}
\begin{IEEEeqnarray}{rCl}
P_{e} &=&  
\frac{1}{\nu_1 \nu_2 \nu'_2}\sum_{m_1, m_2, m_2'}
 P_{Y_1,Y_2|X}(S_e|f(m_1,m_2,m_2')) \label{eq:prob_error_def_1}\\
P'_{e} &=& \frac{1}{\nu_1 \nu_2 \nu'_2}\sum_{m_1, m_2, m_2'}
 P_{Y_1,Y_2|X}(S'_e|f(m_1,m_2,m_2')) \label{eq:prob_error_def_prime}
\end{IEEEeqnarray}
\end{subequations}
where the sets $S_e$ and $S'_e$ are defined as
\begin{align}
&S_e (m_1,m_2) =
 \left\{(\bml{y}_1,\bml{y}_2): g_1(\bml{y}_1)\not=m_1\mbox{ or }
g_2(\bml{y}_2)\not=m_2\right\},\nonumber\\
&S'_e(m_1,m_2,m_2') =
S_e(m_1,m_2)\nonumber\\
&\hspace{2.7cm}\cup\left\{(\bml{y_1},\bml{y}_2):g_2'(\bml{y}_2,h(\bml{y}_1))\not=m_2'\right\},\label{threeEvents} 
\end{align}
and for notational convenience, the dependence of $S_e$ and $S'_e$
on the messages is dropped in~\eqref{eq:prob_error_def}.
\end{definition}

The conference rate $C_{1,2}$ and the communications rates $(R_1,R_2,R_2')$ are defined
as usual:
\[
C_{1,2} = \frac{\log\nu_{1,2}}{n},\quad 
R_k=\frac{\log\nu_k}{n}, k=1,2,\quad R_2'=\frac{\log\nu_2'}{n}.
\]
The interpretation of the rates is as follows: $C_{1,2}$ is the conference
rate in case that it is present. The rate $R_k$ is intended to User $k$, $k=1,2$,
to be decoded whether or not the conference is present. The rate $R_2'$ is intended to User 2
and is the extra rate gained if the conference link is present. 

A rate quadruple $(R_1,R_2,R'_2,C_{1,2})$ is said to be achievable with
unreliable conference if for any  $\epsilon>0$, $\gamma>0$, and sufficiently large $n$
there exists an $(n,e^{n(R_1-\gamma)},e^{n(R_2-\gamma)},e^{n(R_2'-\gamma)},e^{n(C_{1,2}+\gamma)},\epsilon)$
code for the BC with unreliable conference link. The capacity region is the closure of the
set of all achievable quadruples $(R_1,R_2,R'_2,C_{1,2})$ and is denoted by ${\cal C}$. 
For a given conference rate $C_{1,2}$, ${\cal C}(C_{1,2})$ stands for the section
of ${\cal C}$ at $C_{1,2}$. Our interest is to characterize ${\cal C}(C_{1,2})$.

Let  ${\cal R}(C_{1,2})$ be the convex hull of all rate triples $(R_1,R_2,R_2')$ satisfying:
\begin{subequations}
\label{eq:regionBC}
\begin{IEEEeqnarray}{rCl}
R_1 &\leq& I(X;Y_1|U,V),\label{eq:regionBC_R1}\\
R_2 &\leq& I(U;Y_2),\label{eq:regionBC_R2}\\
R_2' &\leq& \min\left\{I(V;Y_2|U) + C_{1,2}, I(V;Y_1|U)\right\}, \label{eq:regionBC_R2prime}
\end{IEEEeqnarray}
for some joint distribution of the form
\begin{equation}
P_{U,V,X,Y_1,Y_2} = P_{U,V}P_{X|U,V}P_{Y_1,Y_2|X},\label{eq:regionBC_joint}
\end{equation}
where $\abs{\calU}\leq\abs{\calX}+3$, and $\abs{\calV}\leq(\abs{\calX}+2)(\abs{\calX}+3)$. 
\end{subequations}
Our main result on the physical degraded BC with unreliable conference is the following
\begin{theorem}
\label{theo:BC}
For any physically degraded BC with unreliable conference of rate $C_{1,2}$,
\[
{\cal C}(C_{1,2}) = {\cal R}(C_{1,2}).
\]
\end{theorem}
The proof is given in Section~\ref{sec:proofs}. Given the last result, we make the following observations:
\begin{itemize}[leftmargin=*]
\item The direct part in the proof of Theorem~\ref{theo:BC} is based on a combination of superposition coding and binning. The intuitive explanation/interpretation of the various auxiliary random variables in \eqref{eq:regionBC} is as follows. First, the information of User 2 is encoded with $U$, which depends on the message $M_2$. This information is always decoded by both decoders, whether the conference link is present or not. The extra message of User 2, which is $M_2'$ is encoded with $V$, which is superimposed on top of $U$. Finally, the message of User 1, namely $M_1$ is encoded with $X$, which is again superimposed on top of $U$ and $V$. The information encoded with $U$, $V$, and $X$, are always decoded by the first decoder, whether the conference link is present or not. The extra information encoded with $V$, however, is decoded (with the help of the conference link) by the second decoder only if the conference link is present. This is done by using the binning approach.
\item Let us examine the region ${\cal R}(C_{1,2})$ when $C_{1,2}=0$, that is, the case where even if the conference link
is present, its rate is 0, and there is no benefit from the conference link. Due to~\eqref{eq:regionBC_joint} the Markov condition $(U,V)\markov Y_1\markov Y_2$ holds, implying, of course, also that $V\markov(U,Y_1)\markov Y_2$ holds. Therefore, when $C_{1,2}=0$, it is readily seen that the bounds in~\eqref{eq:regionBC} reduce to
\begin{subequations}
\label{eq:regionBC_0conf}
\begin{IEEEeqnarray}{rCl}
R_1 &\leq& I(X;Y_1|U,V). \label{eq:regionBC_R1_0conf}\\
R_2 &\leq& I(U;Y_2),\label{eq:regionBC_R2_0conf}\\
R_2' &\leq& I(V;Y_2|U), \label{eq:regionBC_R2prime_0conf}
\end{IEEEeqnarray}
\end{subequations}
The total rate to User 2 is $R_2+R_2'$. Now, it is easy to verify that
after optimization over $(U,V)$, the rates guaranteed by~\eqref{eq:regionBC_0conf}
coincide with the capacity region of the degraded BC, as one should expect. Indeed, we have:
\begin{subequations}
\begin{IEEEeqnarray}{rCl}
R_1 &\leq& I(X;Y_1|U,V),\\
R_2'+R_2 &\leq& I(U,V;Y_2),
\end{IEEEeqnarray}%
\end{subequations}
and so, by letting $\tilde{U}\triangleq (U,V)$ where $P_{\tilde{U},X,Y_1,Y_2} = P_{\tilde{U}}P_{X|\tilde{U}}P_{Y_1,Y_2|X}$, we obtain the capacity region of the degraded BC. 
\item Another case of interest is when $R_2=0$. 
Here, User 2 will not get any rate if the conference link is absent. Choosing $U$
to be a null RV, the region of rates $(R_1,R_2')$
guaranteed by~\eqref{eq:regionBC} reduces to
\begin{subequations}\label{Daboraregion}
\begin{IEEEeqnarray}{rCl}
R_1 &\leq& I(X;Y_1|V),\\
R_2' &\leq& \min\left\{I(V;Y_2) + C_{1,2}, I(V;Y_1)\right\},
\end{IEEEeqnarray}%
\end{subequations}
which coincides with the result in~\cite[Theorem 1]{J_DaboraServetto06Broadcast}.
\item Theorem \ref{theo:BC} can be easily generalized to encounter cases in which there is an input constraint of the form $\bE\pp{\sum_{i=1}^n\Gamma(X_i)}\leq nP$. In this case the achievable region is given by Theorem \ref{theo:BC} where the additional constraint $\bE\pp{\Gamma(X_1)}\leq P$ is needed. Note that the achievability and the converse proofs of Theorem \ref{theo:BC} with the input constraint remain the same, where in the achievability part, we make use of the fact that by the law of large numbers the constraints are satisfied with high probability.
\item It is interesting to check what happens in case that the rate to User 2 is smaller than the capacity of the cooperation link, namely, $R_2'\leq C_{1,2}$. When the cooperation link is reliable (i.e., always available), which is the model considered in \cite[Theorem 1]{J_DaboraServetto06Broadcast}, it can be shown that the capacity region is the convex hull of all rate pairs $(R_1,R_2')$ such that
\begin{subequations}\label{DaboraregionC12}
\begin{IEEEeqnarray}{rCl}
R_1+R_2' &\leq& I(X;Y_1),\label{Dag1}\\
R_2' &\leq& C_{1,2}.\label{Dag2}
\end{IEEEeqnarray}%
\end{subequations}
for some joint distribution $P_{X,Y_1} = P_{X}P_{Y_1|X}$. This result is indeed reasonable due to the fact that in this case User 1 can transmit all the information through the cooperation link. To show \eqref{DaboraregionC12}, first note that the intersection between \eqref{Daboraregion} and $R_2'\leq C_{1,2}$ gives
\begin{subequations}\label{Daboraregion2}
\begin{IEEEeqnarray}{rCl}
R_1 &\leq& I(X;Y_1|V),\label{Dag0}\\
R_2' &\leq& \min\left\{C_{1,2}, I(V;Y_1)\right\},\label{Dag01}
\end{IEEEeqnarray}%
\end{subequations}
for some joint distribution $P_{V,X,Y_1} = P_{V}P_{X|V}P_{Y_1|X}$. Let $\mathscr{A}$ and $\mathscr{B}$ denote the regions in \eqref{DaboraregionC12} and \eqref{Daboraregion2}, respectively. Then, it is evident that $\mathscr{B}\subseteq\mathscr{A}$ due to the Markov chain $V\markov X\markov Y_1$. We now proceed to show the reverse inclusion, i.e., $\mathscr{A}\subseteq\mathscr{B}$. To this end, let $(R_1,R_2')\in\mathscr{A}$, achieved by some $X$. We consider two cases: if $R_1=I(X;Y_1)$, then by using \eqref{Dag1}, we get
$R_2'=0$. However, from \eqref{Dag0} we see that $R_1=I(X;Y_1)$ if and only if $V=\emptyset$, from which we also get that $R_2'=0$. Thus, $(R_1,R_2')\in\mathscr{B}$. If, however, $R_1<I(X;Y_1)$, then let $R_1 = I(X;Y_1)-\alpha$, for $\alpha>0$. We define
\begin{align}
V\triangleq\begin{cases}
\emptyset,\ &\text{w.p. }\beta\\
X,\ &\text{w.p. }1-\beta
\end{cases},
\end{align}
for some $\beta\in\left[0,1\right)$. Obviously, we have the Markov chain $V\markov X\markov Y_1$, and it is easy to see that
\begin{align}
I(X;Y_1|V) = \beta\cdot I(X;Y_1).
\end{align}
Now, since $\beta$ is arbitrary, we can choose as
\begin{align}
\beta = \frac{I(X;Y_1)-\alpha}{I(X;Y_1)},
\end{align}
and we readily get that
\begin{align}
R_1 = I(X;Y_1)-\alpha = I(X;Y_1|V),\label{Dag4}
\end{align}
and
\begin{align}
R_1+R_2'\leq I(X;Y_1) = I(V;Y_1).\label{Dag5}
\end{align}
Combining \eqref{Dag2}, \eqref{Dag4}, and \eqref{Dag5}, we have that $(R_1,R_2)$ satisfy
\begin{subequations}\label{Daboraregion2satisfy}
\begin{IEEEeqnarray}{rCl}
R_1 &\leq& I(X;Y_1|V),\\
R_2' &\leq& \min\big\{C_{1,2}, I(V;Y_1)\big\},
\end{IEEEeqnarray}%
\end{subequations}
for $V\markov X\markov Y_1$, which implies that $(R_1,R_2')\in\mathscr{B}$.

\hspace{0.35cm}When the cooperation link is unreliable, however, using the same arguments as above, it can be shown that the capacity region when $R_2'\leq C_{1,2}$ is the convex hull of all rate triples $(R_1,R_2,R_2')$ that satisfy
\begin{subequations}
\label{eq:regionBCSmallerC12}
\begin{IEEEeqnarray}{rCl}
R_1+R_2' &\leq& I(X;Y_1|U),\\
R_2 &\leq& I(U;Y_2),\\
R_2'&\leq& C_{1,2},
\end{IEEEeqnarray}
\end{subequations}
for some distribution $P_{U,X,Y_1,Y_2} = P_{U}P_{X|U}P_{Y_1,Y_2|X}$. This result makes sense because of the fact that when the cooperation link is absent, we still would like to transmit some information to the User 2, which is captured by $U$.
\end{itemize}

To illustrate the general result in Theorem \ref{theo:BC}, we consider the following simple example.
\begin{example}\label{exmp:1}
Consider the example where the channel output $Y_1$ is clean, namely, $Y_1=X\in\ppp{0,1}$, and $Y_2$ is the output of a binary symmetric channel, i.e., $Y_2 = X\oplus Z$, where $Z$ is Bernoulli with $\Pr\ppp{Z=0}=p$ and statistically independent of $X$. In this case, we obtain from Theorem \ref{theo:BC} that the capacity region is:
\begin{subequations}
\label{eq:regionBCExample}
\begin{IEEEeqnarray}{rCl}
R_1 &\leq& H(X|U,V),\\
R_2 &\leq& I(U;Y_2),\\
R_2' &\leq& \min\left\{I(V;Y_2|U)+C_{1,2}, I(V;X|U)\right\}.
\end{IEEEeqnarray}
\end{subequations}
Fig. \ref{fig:exmp0} depicts the capacity region in \eqref{eq:regionBCExample}, assuming that $C_{1,2}=0.5$, for several values of $R_2$. We present four curves corresponding to the capacity region of the standard BC without cooperation (black curve), the rates $(R_1,R_2)$ which refer to the case where (unreliable) conferencing/cooperation is absent (blue dashed curve), the rates $(R_1,R_2+R_2')$ which refer to the case where (unreliable) conferencing/cooperation is present (red doted curve), and the capacity region in case of reliable/perfect cooperation \cite{J_DaboraServetto06Broadcast} (green dashed-doted curve), i.e., regular cooperation with reliable link. It can be seen that (total) higher rates can be achieved in case of unreliable cooperation compared to the case where there is no cooperation at all, as expected. Also, comparing the $(R_1,R_2+R_2')$ curve and the reliable cooperation curve, it can be noticed that there is some degradation due to the fact that the cooperation link is unreliable. Finally, from the $(R_1,R_2)$ and the standard BC curves it can be seen that the there is some price in terms of the rate $R_2$ (namely, when there is no cooperation) due to the universality of the coding scheme in case of unreliable cribbing.

\begin{figure}[!t]
\begin{minipage}[b]{1.0\linewidth}
  \centering
	\centerline{\includegraphics[width=8cm,height = 6cm]{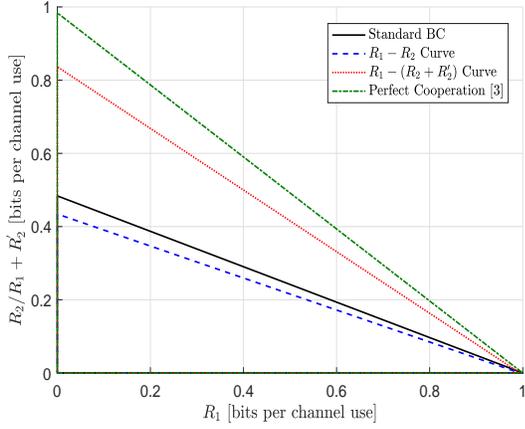}}
	\end{minipage}
\caption{The capacity region of the BC in Example \ref{exmp:1} with $C_{1,2}=0.5$, compared to the capacity region of the BC with reliable cooperation \cite{J_DaboraServetto06Broadcast}.}
\label{fig:exmp0}
\end{figure}
\end{example}

\section{The Multiple Access Channel with Cribbing Encoders}
\label{sec:MAC_cribbing}
A multiple access channel (MAC) is a quadruple 
$({\cal X}_1,{\cal X}_2,{\cal Y},P_{Y|X_1,X_2})$, where ${\cal X}_k$
is the input alphabet of User $k$, $k=1,2$, ${\cal Y}$ is the output alphabet,
and $P_{Y|X_1,X_2}$ is the transition probability matrix
from ${\cal X}_1\times{\cal X}_2$ to ${\cal Y}$. The channel
is memoryless without feedback.

In this section we present achievable rates for 
the MAC with an unreliable cribbing - that may or may not be present -
from Encoder 1 to Encoder 2. The basic assumptions are
as follows. Since Encoder 2 listens to Encoder 1,
he knows whether the cribbing link is present. Similarly, the decoder knows 
it since Encoder 2 can convey to him this message, as it is only one bit of information
to transmit. Encoder 1, on the other hand, does not know
whether the cribbing link is present, since he cannot be informed about it. 
He is only aware that cribbing could occur.
Let ${\cal N}_1'=\left\{1,2,\ldots,\nu_1'\right\}$ and 
${\cal N}_2''=\left\{1,2,\ldots,\nu_2''\right\}$ be two message sets.
A coding scheme
operates as follows. Four messages $M_1$, $M_1'$, $M_2$, and
$M_2''$ are drawn uniformly and independently from the sets
${\cal N}_1$, ${\cal N}_1'$, ${\cal N}_2$, ${\cal N}_2''$, respectively.
Encoder 1 maps the pair $(M_1,M_1')$ to an input sequence
$\bml{x}_1=\bml{x}_1(M_1,M_1')$. If the cribbing link is absent, Encoder 2 maps
the message $M_2$ to to an input sequence 
$\bml{x}_2=\bml{x}_2(M_2)$.  If the cribbing link is present, 
Encoder 2 knows $\bml{x}_1$ strictly causally, thus maps the pair
$(M_2'',\bml{x}_1)$ to an input sequence $\bml{x_2}''$, in a strictly causal manner:
\begin{IEEEeqnarray}{rCl}
\bml{x}_2''(m_2'',\bml{x}_1) &=& (x_{2,1}''(m_2''),x_{2,2}''(m_2'',x_{1,1}),\nonumber\\
                    & & \ \ \ \ \ \ \ \ \ \ \ \ \ \ \ldots,x_{2,n}''(m_2'',x^{n-1}_{1})). \label{eq:MAC_x2}
\end{IEEEeqnarray}
At the output, the decoder decodes $(M_1,M_2)$ if cribbing is absent,
and $(M_1,M_1',M_2'')$ if cribbing is present.

\begin{figure}[!t]\centering
\psscalebox{0.65 0.65}
{
\begin{pspicture}(0,-1.8)(11.74,1.8)
\psframe[linecolor=black, linewidth=0.04, dimen=outer](9.2,0.6)(7.2,-0.2)
\psframe[linecolor=black, linewidth=0.04, dimen=outer](6.0,1.8)(3.6,-1.4)
\psframe[linecolor=black, linewidth=0.04, dimen=outer](2.4,1.8)(0.0,1.0)
\psframe[linecolor=black, linewidth=0.04, dimen=outer](2.4,-0.6)(0.0,-1.4)
\psline[linecolor=black, linewidth=0.04, arrowsize=0.05291666666666668cm 2.0,arrowlength=1.4,arrowinset=0.0]{->}(2.4,1.4)(3.6,1.4)
\psline[linecolor=black, linewidth=0.04, arrowsize=0.05291666666666668cm 2.0,arrowlength=1.4,arrowinset=0.0]{->}(6.0,0.2)(7.2,0.2)
\psline[linecolor=black, linewidth=0.04, arrowsize=0.05291666666666668cm 2.0,arrowlength=1.4,arrowinset=0.0]{->}(9.2,0.2)(10.0,0.2)
\rput[bl](2.9,1.5){$X_1^n$}
\rput[bl](3.85,0.1){$P(Y\vert X_1,X_2)$}
\rput[bl](6.4,0.3){$Y^n$}
\rput[bl](0.4,1.35){Encoder 1}
\rput[bl](0.4,-1.05){Encoder 2}
\rput[bl](9.6,0.35){$(\hat{M}_1,M_2)$}
\rput[bl](2.9,-1.9){Multiple Access Channel}
\psline[linecolor=black, linewidth=0.04, arrowsize=0.05291666666666668cm 2.0,arrowlength=1.4,arrowinset=0.0]{->}(2.4,-1.0)(3.6,-1.0)
\rput[bl](2.9,-0.9){$X_2^n$}
\rput[bl](7.6,0.15){Decoder}
\psline[linecolor=black, linewidth=0.04, dotsize=0.07055555555555555cm 2.0]{*-}(2.8,1.4)(2.8,0.6)
\psline[linecolor=black, linewidth=0.04, dotsize=0.07055555555555555cm 2.0]{-**}(2.8,0.6)(2.4,0.2)
\psline[linecolor=black, linewidth=0.04, dotsize=0.07055555555555555cm 2.0]{o-*}(2.8,0.2)(2.8,-1.0)
\rput[bl](9.6,-0.35){$(\hat{M}_1,\hat{M}'_1,M_2'')$}
\end{pspicture}
}
\caption{MAC with unreliable cribbing encoders.}
\label{fig:2}
\end{figure}
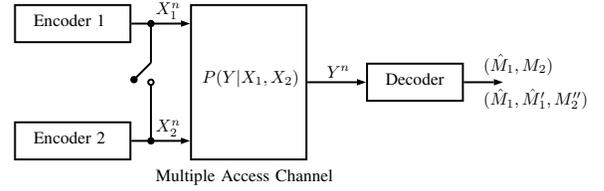

Note that there is a slight difference in the interpretation of the message sets,
compared to the message sets of the BC model studied in Section~\ref{sec:PDBC_cooperating}.
The pair $(M_1,M_1')$ is encoded by User 1, where $M_1$ is always decoded,
and $M_1'$ is decoded only if cribbing is present.
For User 2, if cribbing is absent, $M_2$ is encoded, whereas if cribbing is present, $M_2''$
is encoded. Therefore User 2 can reduce his rate in case of cribbing, in favor
of increasing the rate of User 1. Due to this structure, the joint distribution
of $M_2$ and $M_2''$ is immaterial, as they never appear together in the coding scheme. The setting of the problem is depicted in Fig. \ref{fig:2}.

Following is a formal definition of the scheme described above.
\begin{definition}
\label{def:MAC_code}
An $(n,\nu_1,\nu_1',\nu_2,\nu_2'',\epsilon)$ code for the MAC $P_{Y|X_1,X_2}$
with unreliable strictly causal cribbing link consist of $n+2$ encoding maps
\begin{subequations}
\label{eq:MAC_enc}
\begin{IEEEeqnarray}{rCl}
& & f_1: {\cal N}_1\times{\cal N}_1' \rightarrow {\cal X}_1^n,\label{eq:MAC_enc_1}\\
& & f_2: {\cal N}_2\rightarrow {\cal X}_2^n,\label{eq:MAC_enc_2}\\
& & f_{2,i}'':{\cal N}_2''\times{\cal X}_1^{i-1}\rightarrow {\cal X}_{2,i},\quad i=1,2,\ldots,n,\label{eq:MAC_enc_3}
\end{IEEEeqnarray}
\end{subequations}
and a pair of decoding maps
\begin{subequations}
\label{eq:MAC_dec}
\begin{IEEEeqnarray}{rCl}
& & g: {\cal Y}^n\rightarrow {\cal N}_1\times {\cal N}_2,\label{eq:MAC_dec_1}\\
& & g':{\cal Y}^n\rightarrow {\cal N}_1\times{\cal N}_1'\times{\cal N}_2'',\label{eq:MAC_dec_2}
\end{IEEEeqnarray}
\end{subequations}
such that the average probabilities of error $P_e$ and $P_e'$ do not exceed $\epsilon$. Here
\begin{subequations}
\label{eq:MAC_Pe}
\begin{IEEEeqnarray}{rCl}
P_e &=& \frac{1}{\nu_1\nu_1'\nu_2}\sum_{m_1,m_1',m_2}
P_{Y|X_1,X_2}({\cal Q}_e|f_1(m_1,m_1'),f_2(m_2))\nonumber\\
\label{eq:MAC_Pe_1}\\
P_e'&=& \frac{1}{\nu_1\nu_1'\nu_2''}\sum_{m_1,m_1',m_2''}\nonumber\\
& &\hspace{-0.2cm}P_{Y|X_1,X_2}({\cal Q}_e'|f_1(m_1,m_1'),\bml{f}_2''(m_2'',f_1(m_1,m_1')))\label{eq:MAC_Pe_2}
\end{IEEEeqnarray}
\end{subequations}
where $\bml{f}_2''(m_2'',f_1(m_1,m_1'))$ is the sequence of maps $f_{2,i}''$  in~\eqref{eq:MAC_enc_3},
the sets ${\cal Q}_e$ and ${\cal Q}_e'$ are defined as
\begin{subequations}
\label{depQe}
\begin{IEEEeqnarray}{rCl}
{\cal Q}_e(m_1,m_2) &=& \left\{\bml{y}:g(\bml{y})\not=(m_1,m_2)\right\},\\
{\cal Q}_e'(m_1,m_1',m_2'') &=& \left\{\bml{y}:g'(\bml{y})\not=(m_1,m_1',m_2'')\right\},
\end{IEEEeqnarray}
\end{subequations}
and the dependence of the sets ${\cal Q}_e$, ${\cal Q}_e'$ on the messages
is dropped in~\eqref{eq:MAC_Pe}, for notational convenience.
\end{definition}

The rates $(R_1,R_1',R_2,R_2'')$, and achievability of a given quadruple, are defined as usual.
The capacity region of the MAC with unreliable strictly causal cribbing is the closure of the collection
of all achievable quadruples $(R_1,R_1',R_2,R_2'')$, and is denoted by ${\cal C}_{\text{mac}}^{\text{strict}}$.
Our interest is in characterizing ${\cal C}_{\text{mac}}^{\text{strict}}$.

Let $\calU$ and ${\cal V}$, be finite sets, and let ${\cal P}^{\text{strict}}$ be the collection of all joint distributions
$P_{U,V,X_1,X_2,X_2'',Y,Y''}$ of the form
\begin{align}
P_{U}P_{V}P_{X_1|U,V}P_{X_2}P_{Y|X_1,X_2} P_{X_2''|U} P_{Y''|X_1,X_2''}\label{causregremdis}
\end{align}
where $P_{Y''|X_1,X_2''}$ is our MAC with $X_2''$ at the input of Encoder 2. Let ${\cal I}^{\text{strict}}_{\text{mac}}$ be the convex hull of all rate quadruples $(R_1,R_1',R_2,R_2'')$ satisfying
\begin{subequations}\label{causregrem}
\begin{IEEEeqnarray}{rCl}
R_1 &\leq& I(V;Y|X_2),\\
R_2 &\leq& I(X_2;Y|V),\\
R_1+R_2 &\leq& I(V,X_2;Y),\\
R_1'&\leq&H(X_1\vert U,V),\\
R_2''&\leq&I(X_2'';Y''\vert U,V,X_1),\label{19e}\\
R_1'+R_2''&\leq&I(X_1,X_2'';Y''\vert V),\\
R_1+R_1'+R_2''&\leq&I(X_1,X_2'';Y''),
\end{IEEEeqnarray}%
\end{subequations}
for some $P_{U,V,X_1,X_2,X_2'',Y,Y''}\in\calP^{\text{strict}}$ where
\begin{align}
\abs{\calU}&\leq\min\ppp{\abs{\calX_1}\cdot\abs{\calX_2}+1,\abs{\calY}+2}\\
\abs{\calV}&\leq\min\ppp{\abs{\calX_1}\cdot\abs{\calX_2}+4,\abs{\calY}+5}.
\end{align}
We start with the following result, which is proved in Subsection \ref{sub:achMac}.
\begin{theorem}[Inner bound - strictly causal case]
\label{theo:MACstr}
For any MAC with unreliable strictly causal cribbing
\[
{\cal I}_{\text{mac}}^{\text{strict}}\subseteq {\cal C}_{\text{mac}}^{\text{strict}}.
\]
\end{theorem}

Next, consider the case where causal cribbing, for the second user, is allowed, that is,
\begin{align}
&\bml{x}_2''(m_2'',\bml{x}_1) = (x_{2,1}''(m_2'',x_{1,1}),\ldots,x_{2,n}''(m_2'',x^{n}_{1})),\label{eq:MAC_enc_3str}
\end{align}	
or, equivalently, replace \eqref{eq:MAC_enc_3} with:
\begin{align}								
&f_{2,i}'':{\cal N}_2''\times{\cal X}_1^{i}\rightarrow {\cal X}_{2,i},\quad i=1,2,\ldots,n.\label{eq:MAC_x2str}
\end{align}
The capacity ${\cal C}_{\text{mac}}$ of the MAC with unreliable causal cribbing is defined similarly to the strictly causal case, but with \eqref{eq:MAC_enc_3str} and \eqref{eq:MAC_x2str}, replacing \eqref{eq:MAC_x2} and \eqref{eq:MAC_enc_3}, respectively.

Let ${\cal P}$ be the collection of all joint distributions
$P_{V,X_1,X_2,X_2'',Y,Y''}$ of the form
\begin{IEEEeqnarray}{rCl}
P_{V,X_1}P_{X_2}P_{Y|X_1,X_2} P_{X_2''|X_1} P_{Y''|X_1,X_2''}. \label{eq:MAC_joint}
\end{IEEEeqnarray}
The interpretation of the coding random variables and their joint distribution is as follows. The pair $(V,X_1)$ are the coding RVs of User 1. These are fixed, regardless of whether cribbing
is present or not. The input $X_2$ is the coding variable of User 2 if cribbing
is absent, therefore it is independent of $(V,X_1)$, and $Y$ is the MAC output
due to inputs $X_1,X_2$. When cribbing is present, User 2 
encodes with $X_2''$ which can depend on $X_1$.
The output of the channel
due to inputs $X_1$ and $X_2''$ is denoted by $Y''$. 

Let ${\cal I}_{\text{mac}}$ be the convex hull of all rate quadruples $(R_1,R_1',R_2,R_2'')$ satisfying
\begin{subequations}
\begin{IEEEeqnarray}{rCl}
R_1 &\leq& I(V;Y|X_2), \label{eq:MAC_R1}\\
R_2 &\leq& I(X_2;Y|V),\label{eq:MAC_R2}\\
R_1+R_2 &\leq& I(V,X_2;Y),\label{eq:MAC_sum}\\
R_1' &\leq& H(X_1|V),\label{eq:MAC_R1_prime}\\
R_2''  &\leq& I(X_2'';Y''|V,X_1),\label{eq:MAC_R2_dprime}\\
R_1'+R_2''&\leq&I(X_1,X_2'';Y''\vert V),\label{eq:MAC_R2_dprime0}\\
R_1+R_1'+R_2''&\leq& I(X_1,X_2'';Y''),\label{eq:MAC_sum_prime}
\end{IEEEeqnarray}\label{AchCrib}
\end{subequations}
for some $P_{V,X_1,X_2,X_2'',Y,Y''}\in{\cal P}$ where
\begin{align}
\abs{\calV}\leq\min\ppp{\abs{\calX_1}\cdot\abs{\calX_2}+4,\abs{\calY}+5}.
\end{align}
We have the following result, proved in Subsection \ref{sub:achMac2}.
\begin{theorem}[Inner bound - causal cribbing]
\label{theo:MAC1}
For any MAC with unreliable causal cribbing
\[
{\cal I}_{\text{mac}}\subseteq {\cal C}_{\text{mac}}.
\]
\end{theorem}
We shall make several remarks on the above results.
\begin{itemize}
\item The bounds on the cardinalities of $U$, and $V$, are derived in a similar manner as in \cite[Appendix B]{J_WillemsMeulen85Discrete}, and is based on Fenchel-Eggleston-Carath\'eodry Theorem. 
\item The proof of Theorem~\ref{theo:MACstr} is based on the combination of superposition coding and block-Markov coding. The transmission is always performed in $B$ sub-blocks, of length $n$ each. In each sub-block, the messages of User 1 are encoded in two layers. First, the ``resolution" information of User 1 is encoded with $U$, which depend on both messages $M_1$ and $M_1'$. Then, the fresh information of message $M_1$ is encoded with $V$, and finally, the fresh information of $M_1'$ is encoded with $X_1$, using superposition coding around the cloud centers $V$ and $U$. If the cribbing link is absent, Encoder 2 encodes his messages independently of Encoder 1. The decoder can then decode only the messages of $V$, that is, $M_1$, and $X_2$. If the cribbing link is present, block Markov coding is employed, similarly to the scheme used in~\cite{J_WillemsMeulen85Discrete} for one sided causal cribbing. In this case, the decoder decodes the messages of $V$, $U$, $X_1$, and $X_2''$. Finally, to prove Theorem~\ref{theo:MAC1} we employ Shannon strategies.
\item Note that the main important observation in the achievability, is that User 1 must employ a universal encoding scheme, in the sense of being independent of the cribbing. User 2 and the decoder, however, can employ different encoding and decoding schemes, in accordance to existence or absence of the cribbing.
\item When cribbing is absent, the rates $R_1'$ and $R''_2$ are not decoded. Thus, setting $V=X_1$ in the region ${\cal I}_{\text{mac}}$ yields the capacity region of the MAC without cribbing, as expected.
\item The r.h.s. of \eqref{19e} is smaller than that of \eqref{eq:MAC_R2_dprime}. Indeed,
\begin{align}
I(X_2'';Y''\vert U,V,X_1) &= H(Y''\vert U,V,X_1)\nonumber\\
&\ \ \ - H(Y''\vert U,V,X_1,X_2'')\nonumber\\
&\leq H(Y''\vert V,X_1)- H(Y''\vert V,X_1,X_2'')\nonumber\\
& = I(X_2'';Y''\vert V,X_1)
\end{align}
where the inequality follows from the fact that conditioning reduce entropy, and the Markov chain $(U,V)\markov(X_1,X_2'')\markov Y''$.
\end{itemize}

Unfortunately, we were not able to show the converse part in general, but only for some special case, described in the sequel. In the following, we provide first an outer bound to the capacity region, assuming unreliable causal cribbing.  
Let ${\cal I}_{\text{mac}}^{O}$ be the convex hull of all rate quadruples $(R_1,R_1',R_2,R_2'')$ satisfying
\begin{subequations}\label{AchCribou}
\begin{IEEEeqnarray}{rCl}
R_1 &\leq& I(V;Y|X_2), \label{eq:MAC_R1ou}\\
R_2 &\leq& I(X_2;Y|X_1),\label{eq:MAC_R2ou}\\
R_1+R_2 &\leq& I(V,X_2;Y),\label{eq:MAC_sumou}\\
R_1' &\leq& H(X_1|V),\label{eq:MAC_R1_primeou}\\
R_2''  &\leq& I(X_2'';Y''|V,X_1),\label{eq:MAC_R2_dprimeou}\\
R_1+R_1'+R_2''&\leq& I(X_1,X_2'';Y''),\label{eq:MAC_sum_primeou}
\end{IEEEeqnarray}
\end{subequations}
for some $P_{V,X_1,X_2,X_2'',Y,Y''}\in\calP^{O}$ of the form
\begin{align}
P_{V,X_1}P_{X_2}P_{Y|X_1,X_2} P_{X_2''|V,X_1} P_{Y''|X_1,X_2''}.\label{causregremdisOuter}
\end{align}
The forthcoming result is true also for the non-causal cribbing case, namely,
\begin{align}
&\bml{x}_2''(m_2'',\bml{x}_1) = (x_{2,1}''(m_2'',x_{1}^n),\ldots,x_{2,n}''(m_2'',x^{n}_{1})),\label{eq:MAC_enc_3strnon}
\end{align}	
or, equivalently, replace \eqref{eq:MAC_enc_3} with:
\begin{align}								
&f_{2,i}'':{\cal N}_2''\times{\cal X}_1^{n}\rightarrow {\cal X}_{2,i},\quad i=1,2,\ldots,n.\label{eq:MAC_x2strnon}
\end{align}
The following is proved in Subsection \ref{sub:achMacout}. 
\begin{theorem}[Outer bound - causal (non-causal) case]
\label{theo:MAC1out}
For any MAC with unreliable causal (non-causal) cribbing
\[
{\cal I}_{\text{mac}}^{O}\supseteq {\cal C}_{\text{mac}}.
\]
\end{theorem}
Next, we consider a case in which we were able to derive the capacity region.

Assume that $R_1'=0$, which means that there is no extra rate sent by User 1 to be decoded when cribbing is present. In this case, the first user is fully decoded no matter whether cribbing is present or not. Then, according to Theorem \ref{theo:MAC1}, it is easy to verify that an achievable region is given by:
\begin{subequations}\label{AchCrib2}
\begin{IEEEeqnarray}{rCl}
R_1 &\leq& I(V;Y|X_2), \label{eq:MAC_R101}\\
R_2 &\leq& I(X_2;Y|V),\label{eq:MAC_R201}\\
R_1+R_2 &\leq& I(V,X_2;Y),\label{eq:MAC_sum01}\\
R_2'' &\leq& I(X_2'';Y''\vert V,X_1),\label{eq:MAC_R1_prime01}\\
R_1+R_2''&\leq& I(X_1,X_2'';Y''),\label{eq:MAC_sum_prime01}
\end{IEEEeqnarray}
\end{subequations}
for some $P_{V,X_1,X_2,X_2'',Y,Y''}\in\calP$ of the form
\begin{IEEEeqnarray}{rCl}
P_{V,X_1}P_{X_2}P_{Y|X_1,X_2} P_{X_2''|X_1} P_{Y''|X_1,X_2''}. \label{eq:MAC_jointdist}
\end{IEEEeqnarray}
Let ${\hat{\cal I}}_{\text{mac}}^{I}$ be the convex hull of all rate triples $(R_1,R_2,R_2'')$ satisfying \eqref{AchCrib2} and \eqref{eq:MAC_jointdist}. Next, let ${\hat{\cal I}}_{\text{mac}}^{O}$ be the convex hull of all rate triples $(R_1,R_2,R_2'')$ satisfying:
\begin{subequations}\label{AchCribou2}
\begin{IEEEeqnarray}{rCl}
R_1 &\leq& I(V;Y|X_2),\\
R_2 &\leq& I(X_2;Y|X_1),\\
R_1+R_2 &\leq& I(V,X_2;Y),\\
R_2''  &\leq& I(X_2'';Y''|V,X_1),\\
R_1+R_2''&\leq& I(X_1,X_2'';Y''),
\end{IEEEeqnarray}
\end{subequations}
for some $P_{V,X_1,X_2,X_2'',Y,Y''}\in\calP^O$ in \eqref{causregremdisOuter}. It is easy to see that ${\hat{\cal I}}_{\text{mac}}^{O}$ is obtained upon substitution of $R_1'=0$ in \eqref{AchCribou}, and thus according to Theorem \ref{theo:MAC1out}, it is an outer bound on \eqref{AchCrib2}. In this stage, one may realize that for $R_1'=0$, the auxiliary RV $V$ should be superfluous, and we can actually substitute $X_1$ instead. This is indeed reasonable due to the fact that $V$ is used to convey the message $M_1$, and the extra messages from the first user, that is $M_1'$, is encoded by $X_1$. Accordingly, let ${{\hat{\calI}}}_{\text{mac}}$ be the convex hull of all rate triples $(R_1,R_2,R_2'')$ satisfying:
\begin{subequations}\label{AchCrib3}
\begin{IEEEeqnarray}{rCl}
R_1 &\leq& I(X_1;Y|X_2), \label{eq:MAC_R103}\\
R_2 &\leq& I(X_2;Y|X_1),\label{eq:MAC_R203}\\
R_1+R_2 &\leq& I(X_1,X_2;Y),\label{eq:MAC_sum03}\\
R_2'' &\leq& I(X_2'';Y''\vert X_1),\label{eq:MAC_R1_prime03}\\
R_1+R_2''&\leq& I(X_1,X_2'';Y''),\label{eq:MAC_sum_prime03}
\end{IEEEeqnarray}
\end{subequations}
for some $P_{X_1,X_2,X_2'',Y,Y''}$ of the form
\begin{IEEEeqnarray}{rCl}
P_{X_1}P_{X_2}P_{Y|X_1,X_2} P_{X_2''|X_1} P_{Y''|X_1,X_2''}.\label{AchCrib3Dist}
\end{IEEEeqnarray}
The subsequent lemma is proved in Appendix \ref{app:2}.
\begin{lemma}\label{lem1p} Let ${\hat{\cal I}}_{\text{mac}}^{I}$, ${\hat{\cal I}}_{\text{mac}}^{O}$, and ${{\hat{\calI}}}_{\text{mac}}$, be defined in \eqref{AchCrib2}, \eqref{AchCribou2}, and \eqref{AchCrib3}, respectively. Then,
\begin{align}
{\hat{\cal I}}_{\text{mac}}^{I}= {\hat{\cal I}}_{\text{mac}}^{O} = {{\hat{\calI}}}_{\text{mac}}.
\end{align}
\end{lemma}
Hence, using Lemma \ref{lem1p}, we obtain the following result.
\begin{theorem}
\label{theo:MAC6}
For any MAC with unreliable causal (non-causal) cribbing, if $R_1'=0$, then ${{\hat{\calI}}}_{\text{mac}}$ is the capacity region.
\end{theorem}

According to \eqref{AchCrib3}, if the first user is fully decoded no matter whether cribbing is present or not, then there is no bound on the individual rate of the first user when cribbing is present (we have only bounds on the rate of the second user \eqref{eq:MAC_R1_prime03} and on the sum rate \eqref{eq:MAC_sum_prime03}). Instead, as can be seen from \eqref{eq:MAC_R1_prime03}-\eqref{eq:MAC_sum_prime03}, it is assumed that $X_1$ is already known to the receiver when cribbing is present. The reason is that since cribbing can only help in recovering $X_1$, the bound on the individual rate of the first user when cribbing is absent dominates (or, more strict). 

An interesting conclusion arises from the region in \eqref{AchCrib3}-\eqref{AchCrib3Dist}. Note that \eqref{eq:MAC_R103}-\eqref{eq:MAC_sum03}, when evaluated over all product distributions $P_{X_1}P_{X_2}$ (as in \eqref{AchCrib3Dist}), coincides with the capacity region of the standard MAC, without cribbing. Therefore, for the case of $R_1'=0$, there is no loss of performance when using a robust coding scheme, relative to the case of no cribbing at all. To illustrate the results in Theorems \ref{theo:MAC1} and \ref{theo:MAC6}, and the above conclusion, we consider the following examples. 

\begin{example}\label{exmp:ex2}
Consider the channel model depicted in Fig. \ref{fig:channel}, where the channel output is $Y=(Y_1,Y_2)$, $\rho_0 \triangleq \Pr\ppp{Y_2=1|X_2=0}=\Pr\ppp{Y_2=0|X_2=1}$ and $\rho_1 \triangleq \Pr\ppp{Y_2=1|X_2=2}=\Pr\ppp{Y_2=0|X_2=3}$. The crossover probabilities $\rho_0$ and $\rho_1$ depend on $X_1$ in the following way: if $X_1=0$, then $(\rho_0,\rho_1)=(0,1/2)$, and if $X_1=1$, then $(\rho_0,\rho_1)=(1/2,0)$. Accordingly, if cribbing is present, then User 2 can transmit his information via a noiseless channel. In this case, the total rate that User 2 can transmit is 1, which is the maximal possible rate for him since the output $Y_2$ is binary. When cribbing is absent, User 2 cannot know which of the channels is clean, and thus cannot transmit at the maximal rate 1. Let $P_{X_1} \triangleq \Pr\ppp{X_1=0}$, and $p_i\triangleq \Pr\ppp{X_2=i}$, for $i=0,1,2,3$. Using \eqref{AchCrib3}, it is a simple exercise to check that
\begin{subequations}\label{BECExample01}
\begin{IEEEeqnarray}{rCl}
R_1&\leq& \calH_2(P_{X_1}),\label{BECExample1}\\
R_2&\leq& P_{X_1}\calH_2(p_0+\frac{p_2+p_3}{2})+\bar{P}_{X_1}\calH_2(p_2+\frac{p_0+p_1}{2})\nonumber\\
&&-P_{X_1}\cdot(p_2+p_3)-\bar{P}_{X_1}\cdot(p_0+p_1),\label{BECExample2}\\
R_2''&\leq& 1,\label{BECExample3}
\end{IEEEeqnarray}
\end{subequations}
where $\calH_2\p{\cdot}$ is the binary entropy, and note that in this example the sum-rate constraints in \eqref{eq:MAC_sum03} and \eqref{eq:MAC_sum_prime03}, are redundant because they are given by the sum of the individual rate constraints. Also, note that the optimal distribution $P_{X_2''|X_1}$ in this case, is given by
\begin{align}
P_{X_2''|X_1=0} = \begin{cases}
0.5,\ &\text{if }\;X_2''=0\\
0.5,\ &\text{if }\;X_2''=1\\
0,\ &\text{if }\;X_2''=2\\
0,\ &\text{if }\;X_2''=3
\end{cases},\label{X2primedis}
\end{align}
and
\begin{align}
P_{X_2''|X_1=1} = \begin{cases}
0,\ &\text{if }\;X_2''=0\\
0,\ &\text{if }\;X_2''=1\\
0.5,\ &\text{if }\;X_2''=2\\
0.5,\ &\text{if }\;X_2''=3
\end{cases}.\label{X2primedis2}
\end{align}
Fig. \ref{fig:exmp01} presents three curves corresponding to the capacity region of the standard MAC without cribbing (green ``$+$" curve), the rates $(R_1,R_2)$ which refer to the case where cribbing is absent (blue doted curve), and the rates $(R_1,R_2'')$ which refer to the case where cribbing is present (red curve). Each value of $R_1$ is associated with two rates $R_2$ and $R_2''$. For example, if $R_1 = 0.7$ then $R_2\approx 0.8$, $R_2''=1$. It is evident that higher rates can be achieved for the second user due to the cribbing, as expected. Also, it can be seen that the $(R_1,R_2)$ curve coincide with the capacity region of the standard MAC without cribbing, as expected from \eqref{AchCrib3}. This means that the coding scheme for the case of unreliable cribbing is robust for the case of $R_1'=0$. That is, when $R_1'=0$, the uncertainty about the cribbing link does not have negative influence on the performance compared to the case of no cribbing at all. Finally, we mention that in this example, the capacity region in case of reliable cribbing \cite{J_WillemsMeulen85Discrete} coincides with the $(R_1,R_2'')$ curve.

\begin{figure}[!t]
\begin{minipage}[b]{1.0\linewidth}
  \centering
	\centerline{\includegraphics[width=7cm,height = 3cm]{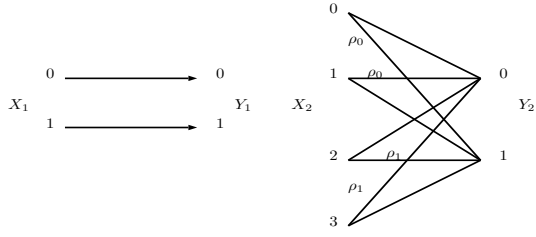}}
	\end{minipage}
\caption{The channel model in Example 1.}
\label{fig:channel}
\end{figure}

\begin{figure}[!t]
\begin{minipage}[b]{1.0\linewidth}
  \centering
	\centerline{\includegraphics[width=8cm,height = 6cm]{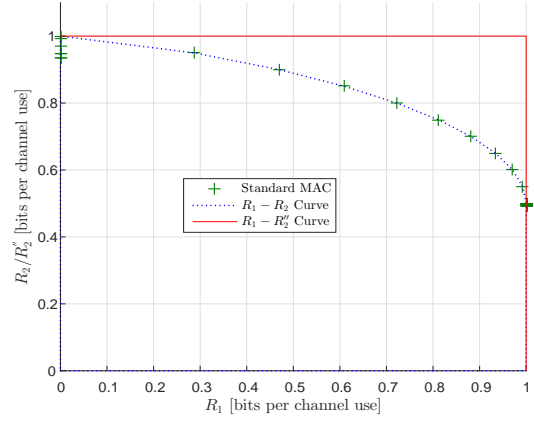}}
	\end{minipage}
\caption{The capacity region of Example~\ref{exmp:ex2} (\eqref{BECExample1}-\eqref{BECExample3}), compared to the capacity region of the standard MAC without cribbing. Each value of $R_1$ is associated with two rates $R_2$ and $R_2''$. The capacity region in case of reliable cribbing coincides with the $(R_1,R_2'')$ curve.}
\label{fig:exmp01}
\end{figure}

\end{example}

\begin{example}\label{exmp:ex3}
Consider the model in the previous example, but now $Y_1 = X_1\oplus Z$, where $Z$ is a Bernoulli RV, independent of $X_1$, and $\Pr\ppp{Z=0}=q$. The capacity region in this case is given in Appendix \ref{app:3} (see \eqref{exmp3cap}). Fig. \ref{fig:exmp012} presents four curves corresponding to the capacity region of the standard MAC without cribbing (green ``$+$" curve), the rates $(R_1,R_2)$ (blue doted curve), the rates $(R_1,R_2'')$ (red curve),  and the capacity region in case of reliable cribbing \cite{J_WillemsMeulen85Discrete} (black dashed curve). In the simulations we choose $q=0.1$. As before, higher rates can be achieved for the second user due to the cribbing, as expected, and it can be seen that the $(R_1,R_2)$ curve coincide with the capacity region of the standard MAC. Finally, contrary to the previous example, here, there is some degradation compared to the reliable cribbing case.

\begin{figure}[!t]
\begin{minipage}[b]{1.0\linewidth}
  \centering
	\centerline{\includegraphics[width=8cm,height = 6cm]{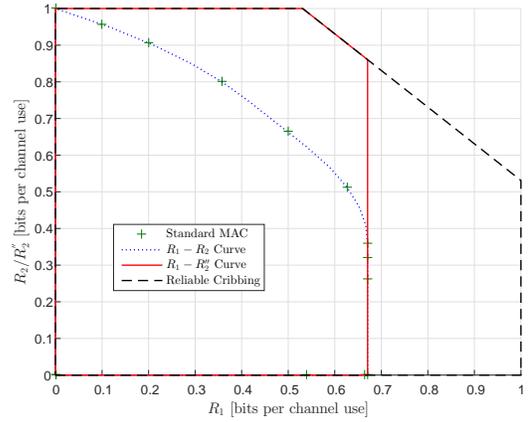}}
	\end{minipage}
\caption{The capacity region in Example \ref{exmp:ex3}, compared to the capacity region of the standard MAC without cribbing, and the capacity region in case of reliable cribbing, for $q=0.1$. Each value of $R_1$ is associated with two rates $R_2$ and $R_2''$.}
\label{fig:exmp012}
\end{figure}

\end{example}

\begin{example}\label{ex:example4}
Consider the example where the channel output, $Y$, is given by:
\begin{align}
Y = X_1\oplus X_2\oplus Z_1\oplus Z_2
\end{align}
where $X_1,X_2,Z_1$, and $Z_2$, are binary RVs, where $Z_1$ is Bernoulli with $\Pr\ppp{Z_1=0}=p_1$, $Z_2=0$ if $X_1=0$, and it is Bernoulli with $\Pr\ppp{Z_2=0}=p_2$, otherwise (i.e., if $X_1=1$). Here, $X_1,X_2,Z_1$ and $Z_2$, are independent. When cribbing is present, the channel output, $Y''$, is given by:
\begin{align}
Y'' = X_1\oplus X_2''\oplus Z_1\oplus Z_2
\end{align}
where now $X_2''$ may depend on $X_1$. Let $\Pr\ppp{X_i=0}\triangleq P_{X_i}$, for $i=1,2$, $\Pr\ppp{X_2''=0\vert X_1=0}=\mu_1$, and $\Pr\ppp{X_2''=0\vert X_1=1}=\mu_2$. Also, for two real numbers $0\leq a,b\leq1$, define $a\ast b \triangleq a\cdot\bar b+\bar a\cdot b$, and $a\star b\triangleq a\cdot b+\bar a\cdot\bar b$, where $\bar a\triangleq 1-a$. Finally, let:
\begin{align}
&\alpha\triangleq (p_1\star p_2)\cdot P_{X_2}+(p_1\ast p_2)\cdot \bar{P}_{X_2},\\
&\beta\triangleq (p_1\star p_2)\cdot \mu_2+(p_1\ast p_2)\cdot \bar\mu_2.
\end{align} 
We wish to evaluate the region in \eqref{AchCrib} where $R_1'$ may be positive. To this end, we choose the auxiliary RV $V$ to be:
\begin{align}
V\triangleq\begin{cases}
\emptyset,\ &\text{w.p. }\; \gamma\\
X_1,\ &\text{w.p. }\; 1-\gamma
\end{cases},
\end{align}
which may be sub-optimal. We define the following quantities:
\begin{align}
I_1&\triangleq \calH_2(P_{X_1}p_1+\bar{P}_{X_1}(p_1\star p_2))-P_{X_1}\calH_2(p_1)\nonumber\\
&\ \ \ \ -\bar{P}_{X_1}\calH_2(p_1\ast p_2),\nonumber\\
I_2&\triangleq \calH_2(P_{X_1}(p_1\star P_{X_2})+\bar P_{X_1}\bar\alpha)-\nonumber\\
&\ \ \ \ -\calH_2(P_{X_1}p_1+\bar{P}_{X_1}(p_1\star p_2)),\nonumber\\
I_3&\triangleq P_{X_1}\calH_2(p_1\star P_{X_2})+\bar P_{X_1}\calH_2(\alpha)-P_{X_1}\calH_2(p_1)\nonumber\\
&\ \ \ \ -\bar{P}_{X_1}\calH_2(p_1\ast p_2),\nonumber\\
I_4&\triangleq \calH_2(P_{X_1}(p_1\star P_{X_2})+\bar P_{X_1}\bar\alpha)-P_{X_1}\calH_2(p_1)\nonumber\\
&\ \ \ \ -\bar{P}_{X_1}\calH_2(p_1\ast p_2),\nonumber\\
I_5&\triangleq P_{X_1}\calH_2(p_1\star\mu_1)+\bar P_{X_1}\calH_2(\beta)-P_{X_1}\calH_2(p_1)\nonumber\\
&\ \ \ \ -\bar{P}_{X_1}\calH_2(p_1\ast p_2),\nonumber\\
I_6&\triangleq \calH_2(P_{X_1}(p_1\star\mu_1)+\bar P_{X_1}\bar\beta)-P_{X_1}\calH_2(p_1)\nonumber\\
&\ \ \ \ -\bar{P}_{X_1}\calH_2(p_1\ast p_2).
\end{align}
Then, using the above definitions, it is a simple exercise to check that \eqref{AchCrib} boils down to:
\begin{subequations}\label{BECExample}
\begin{IEEEeqnarray}{rCl}
R_1&\leq& (1-\gamma)\cdot I_1,\\
R_2&\leq& \gamma\cdot I_2+(1-\gamma)\cdot I_3,\\
R_1+R_2&\leq& \gamma\cdot I_2+(1-\gamma)\cdot I_4,\\
R_1'&\leq& \gamma\cdot\calH_2(P_{X_1}),\\
R_2''&\leq& I_5,\\
R_1'+R_2''&\leq& \gamma\cdot I_6+(1-\gamma)\cdot I_5,\\
R_1+R_1'+R_2''&\leq& I_6.
\end{IEEEeqnarray}
\end{subequations}
Fig. \ref{fig:exmp} depicts the achievable region in \eqref{BECExample} for the case where $p_1=0.01$ and $p_2=0.4$. Since \eqref{BECExample} is parametrized by four rates, it is convenient to fix the rate $R_1'$ on some value, which was chosen in our calculations to be $R_1'=0.3$. We present five curves corresponding to the capacity region of the standard MAC without cribbing (blue ``$+$" curve), the rates $(R_1,R_2)$ (green curve), the rates $(R_1,R_2'')$ (red dashed-doted curve), the rates $(R_1+R_1',R_2'')$ which refer to the total rate of User 1 versus the rate of User 2 when cribbing is present (brown doted curve), and the capacity region in case of reliable cribbing \cite{J_WillemsMeulen85Discrete} (black dashed curve). Each value of $R_1$ is associated with two rates $R_2$ and $R_2''$. For example, if $R_1 = 0.1$ then $R_2\approx 0.75$, $R_2'' \approx 0.56$, and $R_1'=0.3$. This means that when cribbing is present User 2 reduces his rate $R_2''$ in favor of increasing the rate of User $1$. This conclusion is noticeable from the fact that the the $(R_1,R_2)$ curve is on top of the $(R_1,R_2'')$ curve for any $R_1$. Finally, the best results are obtained in the case of reliable cribbing, as expected, and accordingly there is some degradation due to the fact that the cribbing is unreliable.

\begin{figure}[!t]
\begin{minipage}[b]{1.0\linewidth}
  \centering
	\centerline{\includegraphics[width=8cm,height = 6cm]{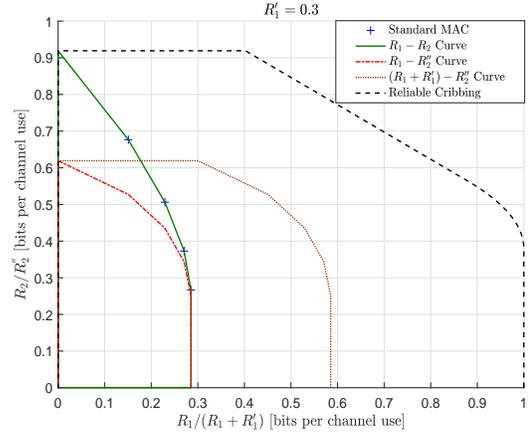}}
	\end{minipage}
\caption{The achievable region in Example~\ref{ex:example4} (see \eqref{BECExample}), for $p_1=0.01$, $p_2=0.4$, and $R_1'=0.3$, compared to the capacity region of the standard MAC without cribbing, and reliable cribbing. Each value of $R_1$ is associated with two rates $R_2$ and $R_2''$ on the $(R_1,R_2)$ and $(R_1,R_2'')$ curves.}
\label{fig:exmp}
\end{figure}

\end{example}

\section{Proofs}
\label{sec:proofs}
\subsection{Proof of Theorem \ref{theo:BC}}
In this subsection, we prove Theorem~\ref{theo:BC}.
The direct part uses random selection and strong typicality arguments.

\bf{\emph{Direct Part}}. \normalfont We start with the code construction.

\emph{Codebook construction:}
Fix a joint distribution $P_{U,V,X}$.
\begin{enumerate}
\item Generate $e^{nR_2}$ codewords $\bml{u}(j)$, $j=1,2,\ldots e^{nR_2}$, i.i.d., according
to $P_U$.
\item For every $\bml{u}(j)$, generate $e^{nR_2'}$ codewords $\bml{v}(k|j)$,
$k=1,2,\ldots e^{nR_2'}$,
independently according to $\prod_{i=1}^n P_{V|U}(v_i|u_i(j))$.
\item For every $j$, distribute  the $e^{nR_2'}$ codewords $\bml{v}(k|j)$,
$k=1,2,\ldots,e^{nR_2'}$, into  $e^{nC_{1,2}}$ bins, evenly and independently
of each other. Thus, in every bin there are $e^{n(R_2'-C_{1,2})}$ codewords 
$\bml{v}(k|j)$ with a fixed index $j$. Denote by $b(k|j)$ the bin number to which
$\bml{v}(k|j)$ belongs. Note that 
\begin{equation}
1\leq b(k|j)\leq e^{nC_{1,2}}.\label{eq:BC_pf_0}
\end{equation}
\item For every pair $(\bml{u}(j),\bml{v}(k|j))$, $j=1,2,\ldots,e^{nR_2}$,
$k=1,2,\ldots,e^{nR_2'}$, generate $e^{nR_1}$ vectors $\bml{x}(l|j,k)$,
$l=1,2,\ldots,e^{nR_1}$, independently of each other, according to
$\prod_{i=1}^nP_{X|U,V}(x_i|u_i(j),v_i(k|j))$.
\end{enumerate}
These codewords form the codebook, which is revealed to the encoder and the decoders.

\emph{Encoding:} Given a triple $(j,k,l)$, where 
$j=1,2,\ldots, e^{nR_2}$, $k=1,2,\ldots, e^{nR_2'}$, $l=1,2,\ldots, e^{nR_1}$, 
the encoder sends via the channel the codeword $\bml{x}(l|j,k)$.

\emph{Decoding:} We assume first that the conference link is absent.
Decoder 2 has $\by_2$ at hand. He looks for the unique index $\hat{j}$
in $\left\{1,2,\ldots,\exp(nR_2)\right\}$ such that
\[
(\bml{u}(\hat{j}),\by_2)\in T^{(n)}_\epsilon(UY_2).
\]
If such $\hat{j}$ does not exist, or there is more than one such index,
an error is declared. By classical results, if
\begin{equation}
R_2 < I(U;Y_2) \label{eq:BC_pf1},
\end{equation}
the index $j$ is decoded correctly with high probability.

Decoder 1 has $\by_1$ at hand. He looks for the unique index
$\hat{\hat{j}}$ in $\left\{1,2,\ldots,\exp(nR_2)\right\}$ such that
\[
(\bml{u}(\hat{\hat{j}}),\by_1)\in T^{(n)}_\epsilon(UY_1).
\]
If such $\hat{\hat{j}}$ does not exist, or there is more than one such index,
an error is declared. By classical results, if
\begin{equation}
R_2 < I(U;Y_1), \label{eq:BC_pf2}
\end{equation}
Decoder 1 succeeds to decode correctly the index $j$ with high probability.
Since the channel is degraded, if~\eqref{eq:BC_pf1} holds, 
it implies~\eqref{eq:BC_pf2}. Next, Decoder 1 looks for the unique index $\hat{\hat{k}}$
in $\left\{1,2,\ldots,\exp(nR_2')\right\}$ such that
\begin{equation}
(\bml{u}(\hat{\hat{j}}),\bml{v}(\hat{\hat{k}}|\hat{\hat{j}}),\by_1) \in T^{(n)}_\epsilon(UVY_1).
\label{eq:BC_pf3}
\end{equation}
If such $\hat{\hat{k}}$ does not exist, or there is more than one such, an error is declared.
By classical results, the index $k\in\left\{1,2,\ldots,\exp(nR_2')\right\}$ is decoded correctly
with high probability if
\begin{equation}
R_2' < I(V;Y_1|U). \label{eq:BC_pf4}
\end{equation}
Having the pair $(\hat{\hat{j}},\hat{\hat{k}})$ at hand, Decoder 1 looks for the unique index 
$\hat{\hat{l}}\in\left\{1,2,\ldots,\exp(nR_1)\right\}$ satisfying
\begin{equation}
(\bml{u}(\hat{\hat{j}}),\bml{v}(\hat{\hat{k}}|\hat{\hat{j}}),\bml{x}(\hat{\hat{l}}|\hat{\hat{k}},\hat{\hat{j}}),
\by_1) \in T^{(n)}_\epsilon(UVXY_1).
\label{eq:BC_pf5}
\end{equation}
By classical results, this step succeeds if the rate $R_1$ satisfies
\begin{equation}
R_1 < I(X;Y_1|U,V). \label{eq:BC_pf6}
\end{equation}
This concludes the decoding process when the conference link is absent. 
By~\eqref{eq:BC_pf1}, \eqref{eq:BC_pf4} and~\eqref{eq:BC_pf6},
the conditions for correct decoding when there is no conferencing
are
\begin{subequations}
\label{eq:BC_pf_region_noconf}
\begin{IEEEeqnarray}{rCl}
R_2 &\leq& I(U;Y_2),\label{eq:BC_pf_region_noconf_R2}\\
R_2' &\leq& I(V;Y_1|U),\label{eq:BC_pf_region_noconf_R2prime}\\
R_1&\leq& I(X;Y_1|U,V).\label{eq:BC_pf_region_noconf_R1}
\end{IEEEeqnarray}
\end{subequations}
Observe that, although the rate $R_2'$ is decoded by Decoder 1 
(if~\eqref{eq:BC_pf_region_noconf_R2prime} is satisfied), it does not arrive to User 2,
since the conferencing link is absent. The bound~\eqref{eq:BC_pf_region_noconf_R2prime}
is still needed in order to guarantee that Decoder 1 can proceed and decode 
the index $l$ (the message intended to him).

We turn now to the case where the conference link is present.
Decoder 1 operates exactly as in the case of no conference, and decodes the indices
$\hat{\hat{j}}$, $\hat{\hat{k}}$, and $\hat{\hat{l}}$. If~\eqref{eq:BC_pf_region_noconf} 
hold, these steps succeed with high probability.
He then sends $b(\hat{\hat{k}}|\hat{\hat{j}})$, the index of the bin
to which $\bml{v}(\hat{\hat{k}}|\hat{\hat{j}})$ belongs, via the conference link.
Due to~\eqref{eq:BC_pf_0}, the link capacity suffices, and Decoder 2 receives
 $b(\hat{\hat{k}}|\hat{\hat{j}})$ without an error.
 
 Decoder 2 decodes the index $\hat{j}$ as in the case of no conference. 
 After receiving from Decoder 1 the bin index $b(\hat{\hat{k}}|\hat{\hat{j}})$,
 he looks in this bin for the unique index $\hat{k}$ such that
 \begin{equation}
 (\bml{u}(\hat{j}),\bml{v}(\hat{k}|\hat{j}),\bml{y}_2)\in T^{(n)}_\epsilon(UVY_2).
 \label{eq:BC_pf_10}
 \end{equation}
 If such an index does not exist, or  there is more than one such,
 an error is declared. From the code construction, every bin contains approximately
 $e^{n(R_2'-C_{1,2})}$ codewords $\bml{v}$. Assuming that the previous
 decoding steps were successful (i.e., $\hat{\hat{j}}$, $\hat{\hat{k}}$, 
 $\hat{j}$ are the correct indices for $j$, $k$, and $j$, respectively),
 by classical results $\hat{k}$  is correct with high probability if
 \begin{equation}
 R_2'-C_{1,2} \leq I(V;Y_2|U). \label{eq:BC_pf_11}
 \end{equation}
 The region defined by~\eqref{eq:BC_pf_region_noconf} and~\eqref{eq:BC_pf_11}
 coincides with ${\cal R}(C_{1,2})$. This concludes the proof of the achievability part.
 
\bf{\emph{Converse Part}}. \normalfont
 We start with a sequence of codes
 $(n,e^{nR_1},e^{nR_2},e^{nR_2'},e^{nC_{1,2}},\epsilon_n)$ with increasing blocklength $n$,
 satisfying $\limn\epsilon_n=0$. 
 We denote by $M_k$ the random message from ${\cal N}_k$, $k=1,2$,
 and by $M_2'$  the message from ${\cal N}_{2}'$. The conference message
 is denoted by $M_{1,2}$.
 By Fano's inequality we can bound the rate $R_2$ as
 \begin{IEEEeqnarray}{rCl}
 nR_2-n\delta_n &\leq& I(M_2;Y_2^n) \\
&\stackrel{(a)}{=}& \sum_{i=1}^nI(M_2;Y_{2,i}|Y_2^{i-1})\nonumber\\
 &\leq& \sum_{i=1}^nI(M_2,Y_2^{i-1};Y_{2,i}),\label{eq:BC_Conv_R2}
 \end{IEEEeqnarray}
 where $\limn\delta_n=0$, due to $\limn\epsilon_n=0$, and (a) follows from the chain rule. 
 We now bound the rate $R_2'$ as follows. If the conference link is present, then
 the messages $M_2'$ can be decoded by Decoder 2 based on $Y_2^n$ and the message
 transmitted via the conference link, $M_{1,2}$. Therefore
 \begin{IEEEeqnarray}{rCl}
 nR_2' - n\delta_n &\leq& I(M_2';Y_2^n,M_{1,2}|M_2)\label{eq:BC_Conv_R2prime}\\
 &=& I(M_2';Y_2^n|M_2) + I(M_2';M_{1,2}|M_2,Y_2^n)\nonumber\\
 &\leq& I(M_2';Y_2^n|M_2) +H(M_{1,2})\nonumber\\
 &=& \sum_{i=1}^n I(M_2';Y_{2,i}|M_2,Y_2^{i-1}) + H(M_{1,2})\nonumber\\
 &\leq& \sum_{i=1}^n I(M_2',Y_1^{i-1};Y_{2,i}|M_2,Y_2^{i-1}) + H(M_{1,2}).\nonumber
 \end{IEEEeqnarray}
 Moreover, the message $M'_{2}$ can be decoded by Decoder 1, regardless of the conference link.
 Hence:
 \begin{IEEEeqnarray}{rCl}
 nR_2'-n\delta_n &\leq& I(M_2';Y_1^n|M_2)\nonumber\\
 &=& \sum_{i=1}^n I(M_2';Y_{1,i}|M_2,Y_1^{i-1})\nonumber\\
 &\stackrel{(a)}{=}& \sum_{i=1}^n I(M_2';Y_{1,i}|M_2,Y_1^{i-1}, Y_2^{i-1})\nonumber\\
 &\leq& \sum_{i=1}^n I(M_2',Y_1^{i-1};Y_{1,i}|M_2,Y_2^{i-1}),\label{eq:BC_Conv_R2prime_2}
 \end{IEEEeqnarray}
 where $(a)$ is true because the channel is physically degraded.
 The rate $R_1$ can be bounded by
 \begin{IEEEeqnarray}{rCl}
 nR_1-n\delta_n &\leq& I(M_1;Y_1^n|M_2,M_2')\nonumber\\
 &=& \sum_{i=1}^n I(M_1;Y_{1,i}|M_2,M_2',Y_1^{i-1})\nonumber\\
 &\stackrel{(a)}{=}& \sum_{i=1}^n I(M_1;Y_{1,i}|M_2,M_2',Y_1^{i-1},Y_2^{i-1})\nonumber\\
 &\stackrel{(b)}{=}& \sum_{i=1}^n I(X_i;Y_{1,i}|M_2,M_2',Y_1^{i-1},Y_2^{i-1}),\label{eq:BC_Conv_R1}
 \end{IEEEeqnarray}
 where $(a)$ is true since the channel is physically degraded. Equality $(b)$ holds since $X_i$
 is a deterministic function of the messages $M_1$, $M_2$, and $M_2'$, and
 since $Y_{1,i}$ is independent of $(M_2,M_2',Y_2^{i-1},Y_1^{i-1},M_1)$ when conditioned
 on $X_i$. Defining $U_i\triangleq(M_2,Y_2^{i-1})$ and $V_i\triangleq(M_2',Y_1^{i-1})$, which due to \eqref{eq:memorylessnofeedback} satisfy the Markov chain $(U_i,V_i)\markov X_i\markov (Y_{1,i},Y_{1,i})$, and using the fact that
 \begin{equation}
 \one H(M_{1,2})\leq C_{1,2}, \label{eq:BC_Conv_M12_bound}
 \end{equation}
 we obtain from~\eqref{eq:BC_Conv_R2}, \eqref{eq:BC_Conv_R2prime}, 
 \eqref{eq:BC_Conv_R2prime_2}, and \eqref{eq:BC_Conv_R1} the bounds
 \begin{subequations}
 \label{eq:BC_Conv_region}
 \begin{IEEEeqnarray}{rCl}
 n(R_2-\delta_n) &\leq& \sum_{i=1}^n I(U_i;Y_{2,i}),\label{eq:BC_Conv_region_R_2}\\
 n(R_2'-\delta_n) &\leq& \sum_{i=1}^n I(V_i;Y_{2,i}|U_i) + nC_{1,2},\label{eq:BC_Conv_region_R2prime_2}\\
 n(R_2'-\delta_n) &\leq& \sum_{i=1}^n I(U_i;Y_{1,i}|V_i), \label{eq:BC_Conv_region_R2prime_1}\\
 n(R_1-\delta_n) &\leq& \sum_{i=1}^n I(X_i;Y_{1,i}|U_i,V_i). \label{eq:BC_Conv_region_R1}
 \end{IEEEeqnarray}
 \end{subequations}
Using the standard time-sharing argument as in \cite[Ch. 14.3]{CoverThomas}, one can rewrite \eqref{eq:BC_Conv_region} by introducing an appropriate time-sharing random variable. Therefore, if $\epsilon_n\to0$ as $n\to\infty$, the convex hull of this region can be shown to be equivalent to the convex hull of the region in \eqref{eq:regionBC}.

Finally, the bounds on the cardinalities of $U$ and $V$ follow from Fenchel-Eggleston-Carath\'eodry Theorem, similarly as used for the 3-receiver degraded BC \cite[Appendix C]{GamalKim}.
\qed

\subsection{Proof of Theorem \ref{theo:MACstr}}\label{sub:achMac}
The proof of Theorem~\ref{theo:MACstr} is based on the combination of superposition coding and block-Markov coding. The transmission is always performed in $B$ sub-blocks, of length $n$ each. In each sub-block, the messages of User 1 are encoded in two layers. First, the ``resolution" information of User 1 are encoded with $U$, which depend on both messages $M_1$ and $M_1'$. Then, the fresh information of message $M_1$ is encoded with $V$, and finally, the fresh information of $M_1'$ is encoded with $X_1$, using superposition coding around the cloud centers $V$ and $U$. If the cribbing link is absent, Encoder 2 encodes his messages independently of Encoder 1. The decoder can then decode only the messages of $V$, that is, $M_1$, and $X_2$. If the cribbing link is present, block Markov coding is employed, similarly to the scheme used in~\cite{J_WillemsMeulen85Discrete} for one sided causal cribbing. 

It is important to emphasize that User 1 must employ a universal encoding scheme, in the sense of being \emph{independent} of the cribbing. User 2 and the decoder, however, can employ different encoding and decoding schemes, in accordance to existence or absence of the cribbing. Accordingly, in the sequel, we describe the encoding scheme for the first user separately. 

We use a random coding argument to demonstrate the achievability part. The messages $M_{1,b}\in\ppp{1,2,\ldots,\exp(nR_1)}$ and $M'_{1,b}\in\ppp{1,2,\ldots,\exp(nR_1')}$, for $b=1,2,\ldots,B-1$, which are uniformly distributed and independent of each other, will be sent over the MAC in $B$ blocks, each of $n$ transmissions. Note that if $B\to\infty$, the overall rates are $R_1(B-1)/B\to R_1$ and $R_1'(B-1)/B\to R_1'$. In each of the $B$ blocks the same codebook is used, and is constructed, for the first user, as follows.

\underline{\emph{Codebook construction for User 1:}} Fix a joint distribution $P_UP_{V}P_{X_1|U,V}$, and a sufficiently small $\epsilon>0$. 
\begin{enumerate}
\item Generate $e^{n(R_1+R_1)}$ codewords $\bml{v}$, i.i.d., according to $P_V$. Label them $\bv(m_0,m_1)$, for $m_0,m_1\in\ppp{1,2,\ldots,\exp(nR_1)}$.
\item Generate $e^{n(R_1+R_1')}$ codewords $\bml{u}$, independently according to $P_U$. Label them $\bml{u}( m_0,m_0')$, for $ m_{0}\in\ppp{1,2,\ldots,\exp(nR_1)}$ and $m_{0}'\in\ppp{1,2,\ldots,\exp(nR_1')}$.
\item For every $\bml{v}(m_0,m_1)$ and $\bml{u}( m_0,m_0')$, generate $e^{nR_1'}$ codewords $\bml{x}_1$, independently according to $\prod_{i=1}^n P_{X_1|U,V}(x_{1,i}|u_i( m_0,m_0'),v_i(m_0,m_1))$. Label them $\bml{x}_1(m_1',\bu( m_0,m_{0}'),\bv(m_0,m_1))$, for $m_{1}'\in\ppp{1,2,\ldots,\exp(nR_1')}$.
\end{enumerate}

We now present the achievability scheme for the case where cribbing is absent.

\bf{\emph{1) Cribbing is absent:}} \normalfont 
The message $M_{2,b}\in\ppp{1,2,\ldots,\exp(nR_2)}$, for $b=1,2,\ldots,B-1$, is uniformly distributed, independent of the messages of the first user, and will be sent over the MAC in $B$ blocks, each of $n$ transmissions. If $B\to\infty$, the overall rate is $R_2(B-1)/B\to R_2$. In each of the $B$ blocks the same codebook is used, and is constructed, for the second user, as follows.

\underline{\emph{Codebook construction for User 2:}} Fix a distribution $P_{X_2}$, and a sufficiently small $\epsilon>0$. Generate $e^{nR_2}$ codewords $\bml{x}_2$, i.i.d., according to $P_{X_2}$. Label them $\bx_2(m_{2})$, for $m_{2}\in\ppp{1,2,\ldots,\exp(nR_2)}$.

The codewords of Users 1 and 2 form the codebook, which is revealed to the encoders and the decoder. The messages $m_{1,b}\in\ppp{1,\ldots,\exp(nR_1)}$, $m'_{1,b}\in\ppp{1,\ldots,\exp(nR_1')}$, and $m_{2,b}\in\ppp{1,\ldots,\exp(nR_2)}$, $b = 1,\ldots,B-1$, are encoded in the following way.

\underline{\emph{Encoding:}} In block 1, the encoders send:
\begin{subequations}
\begin{IEEEeqnarray}{rCl}
\bx_{1,1} &=& \bx_1(m'_{1,1},\bu(1,1),\bv(1,m_{1,1}))\\
\bx_{2,1} &=& \bx_{2}(m_{2,1}).
\end{IEEEeqnarray}
\end{subequations}
Then, in block $b,\;b=2,3,\ldots,B$, the encoders send \eqref{fdfd}, shown at the top of the page.
\begin{figure*}[!t]
\normalsize
\setcounter{MYtempeqncnt}{\value{equation}}
\setcounter{equation}{72}
\begin{subequations}\label{fdfd}
\begin{IEEEeqnarray}{rCl}
\bml{x}_{1,b} &=& \bx_1(m'_{1,b},\bu(m_{1,b-1},m'_{1,b-1}),\bv(m_{1,b-1},m_{1,b})),\ \;b=2,3,\ldots,B-1\\
\bx_{2,b} &=& \bx_{2}(m_{2,b}),\ \;b=2,3,\ldots,B-1\\
\bml{x}_{1B} &=& \bml{x}_1(1,\bu(m_{1,B-1},m'_{1,B-1}),\bv(m_{1,B-1},1)),\\
\bx_{2B} &=& \bx_{2}(m_{2,B}).
\end{IEEEeqnarray}
\end{subequations}
\hrulefill
\vspace*{4pt}
\end{figure*}

\underline{\emph{Decoding:}} We employ simultaneous joint typicality decoding. At the end of the first block, the decoder looks for $(\hat{m}_{1,1},\hat{m}_{2,1})$ such that:
\begin{align}
(\bml{v}(1,\hat{m}_{1,1}),\bml{x}_2(\hat{m}_{2,1}),\by)\in T_\epsilon^{(n)}(VX_2Y).
\end{align}
Next, assume that the decoder has correctly found $\hat{m}_{1,1}$. Then, to find the transmitted information at the end of the second block, the decoder looks for $(\hat{m}_{1,2},\hat{m}_{2,2})$ such that:
\begin{align}
(\bml{v}(\hat{m}_{1,1},\hat{m}_{1,2}),\bml{x}_2(\hat{m}_{2,2}),\by)\in T_\epsilon^{(n)}(VX_2Y).
\end{align}
With the knowledge of $\hat{m}_{1,2}$ the information at the end of the third block can be decoded in a similar manner. In general, at the end of block $b$ the decoder looks for $(\hat{m}_{1,b},\hat{m}_{2,b})$ such that:
\begin{align}
(\bml{v}(\hat{m}_{1,b-1},\hat{m}_{1,b}),\bml{x}_2(\hat{m}_{2,b}),\by)\in T_\epsilon^{(n)}(VX_2Y)
\end{align}
where $\hat{m}_{1,b-1}$ was decoded in the previous block. 

\underline{\emph{Error Analysis:}} By classical results (e.g., standard MAC), there exists a sequence of codes with a probability of error that goes to zero as the block length goes to infinity, if:
\begin{subequations}\label{MAC_R1without}
\begin{IEEEeqnarray}{rCl}
R_1 &\leq& I(V;Y|X_2), \label{eq:MAC_R1_0}\\
R_2 &\leq& I(X_2;Y|V),\label{eq:MAC_R2_1}\\
R_1+R_2 &\leq& I(V,X_2;Y).\label{eq:MAC_sum_2}
\end{IEEEeqnarray}
\end{subequations}
This concludes the decoding process when the conference link is absent.

\bf{\emph{2) Cribbing is present:}} \normalfont 
We turn now to the case where the cribbing link is present. The message $M''_{2,b}\in\ppp{1,2,\ldots,\exp(nR_2'')}$, for $b=1,2,\ldots,B-1$, is uniformly distributed, independent of the messages of the first user, and will be sent over the MAC in $B$ blocks, each of $n$ transmissions. In each of the $B$ blocks the same codebook is used, and is constructed, for the second user, as follows.

\underline{\emph{Codebook construction for User 2:}} Fix a distribution $P_{X_2''|U}$, and a sufficiently small $\epsilon>0$. For every $\bml{u}(m_0,m_0')$, generate $e^{nR_2''}$ codewords $\bml{x}''_2$, independently according to $\prod_{i=1}^n P_{X_2''|U}(x_{2,i}|u_i(m_0,m_0'))$. Label them $\bml{x}_2''(m_2'', \bu(m_0,m_{0}'))$, for $m_{2}''\in\ppp{1,2,\ldots,\exp(nR_2'')}$.
The codewords of Users 1 and 2 form the codebook, which is revealed to the encoders and the decoder.

\underline{\emph{Encoding:}} 
The messages $m_{1,b}\in\ppp{1,\ldots,\exp(nR_1)}$, $m'_{1,b}\in\ppp{1,\ldots,\exp(nR_1')}$, and $m''_{2,b}\in\ppp{1,\ldots,\exp(nR_2'')}$, $b = 1,\ldots,B-1$, are encoded in the following way: In block 1, the encoders send\footnote{Recall that User 1 must employ the same encoding scheme as in the case of absent cribbing.}:
\begin{subequations}
\begin{IEEEeqnarray}{rCl}
\bx_{1,1} &=& \bx_1(m'_{1,1},\bu(1,1),\bv(1,m_{1,1}))\\
\bx_{2,1}'' &=& \bx_{2}''(m''_{2,1},\bu(1,1)).
\end{IEEEeqnarray}
\end{subequations}
Assume that as a result of cribbing from encoder $1$, after block $b,\;b=1,2,\ldots,B-1$, encoder 2 has estimates $\hat{m}_{1,b}$ and $\hat{m}'_{1,b}$, for $m_{1,b}$ and $m'_{1,b}$, respectively. To this end, encoder 2 first chooses $\hat{m}_{1,b}$ such that:
\begin{align}
&(\bv(\hat{m}_{1,b-1},\hat{m}_{1,b}),\bx_{1,b})\in T^{(n)}_\epsilon(VX_1)\label{enc2_1}
\end{align}
where $\hat{m}_{1,b-1}$ was determined at the end of block $b-1$ (recall that ${m}_{1,0}=1$). Then, given $\hat{m}_{1,b}$, he chooses $\hat{m}'_{1,b}$ according to \eqref{enc2_2}, shown at the top of the page, where $\hat{m}'_{1,b-1}$ was determined at the end of block $b-1$. 
\begin{figure*}[!t]
\normalsize
\setcounter{MYtempeqncnt}{\value{equation}}
\setcounter{equation}{79}
\begin{align}
&\left\{\bu(\hat{m}_{1,b-1},\hat{m}'_{1,b-1}),\bv(\hat{m}_{1,b-1},\hat{m}_{1,b}),\right.\left.\bml{x}_1(\hat{m}'_{1,b},\bu(\hat{m}_{1,b-1},\hat{m}'_{1,b-1}),\bv(\hat{m}_{1,b-1},\hat{m}_{1,b})),\bx_{1,b}\right\}\in T^{(n)}_\epsilon(UVX_1X_1)\label{enc2_2}
\end{align}
\hrulefill
\vspace*{4pt}
\end{figure*}
Finally, in block $b,\;b=2,3,\ldots,B$, the encoders send \eqref{fdfd2}, shown at the top of the next page.
\begin{figure*}[!t]
\normalsize
\setcounter{MYtempeqncnt}{\value{equation}}
\setcounter{equation}{80}
\begin{subequations}\label{fdfd2}
\begin{IEEEeqnarray}{rCl}
\bml{x}_{1,b} &=& \bx_1(m'_{1,b},\bu(m_{1,b-1},m'_{1,b-1}),\bv(m_{1,b-1},m_{1,b})),\ \;b=2,3,\ldots,B-1\\
\bx_{2,b}'' &=& \bx_{2}''(m''_{2,b},\bu(\hat{m}_{1,b-1},\hat{m}'_{1,b-1})),\ \;b=2,3,\ldots,B-1\\
\bml{x}_{1B} &=& \bml{x}_1(1,\bu(m_{1,B-1},m'_{1,B-1}),\bv(m_{1,B-1},1)),\\
\bx_{2B}'' &=& \bx_{2}''(1,\bu(\hat{m}_{1,B-1},\hat{m}'_{1,B-1})).
\end{IEEEeqnarray}
\end{subequations}
\hrulefill
\vspace*{4pt}
\end{figure*}

\underline{\emph{Decoding:}} Here, the principle of backward decoding \cite{J_WillemsMeulen85Discrete} is used to find the transmitted information. In the last block, block $B$, the decoder looks for $(\hat{m}_{1,B-1},\hat{m}'_{1,B-1})$ such that 
\begin{align}
&\left\{\bu(\hat{m}_{1,B-1},\hat{m}'_{1,B-1}),\bv(\hat{m}_{1,B-1},1),\right.\nonumber\\
&\hspace{0.5cm}\bml{x}_1(1,\bu(\hat m_{1,B-1},\hat m'_{1,B-1}),\bv(\hat m_{1,B-1},1)),\nonumber\\
&\hspace{0.5cm}\left.\bx_{2}''(1,\bu(\hat m_{1,B-1},\hat m'_{1,B-1})),\by''_B\right\}\nonumber\\
&\hspace{2cm}\in T^{(n)}_\epsilon(UVX_1X_2''Y'').
\end{align}
Next, in block $B-1$, the decoder has at hand an estimate of the fresh information sent in block $B-1$, namely, $(\hat{m}_{1,B-1},\hat{m}'_{1,B-1})$, and to find the transmitted information in block $B-1$ the decoder looks for\footnote{The messages $({m}_{1,B-2},{m}'_{1,B-2})$ are the resolution information of user 1 at block $B-1$, which are actually the fresh messages of $B-2$.} $(\hat{m}_{1,B-2},\hat{m}'_{1,B-2},\hat{m}''_{2,B-1})$ according to \eqref{decodPr}, shown at the top of the page. 
\begin{figure*}[!t]
\normalsize
\setcounter{MYtempeqncnt}{\value{equation}}
\setcounter{equation}{82}
\begin{align}
&\left\{\bu(\hat{m}_{1,B-2},\hat{m}'_{1,B-2}),\bv(\hat{m}_{1,B-2},\hat{m}_{1,B-1}),\bml{x}_1(\hat{m}'_{1,B-1},\bu(\hat m_{1,B-2},\hat m'_{1,B-2}),\bv(\hat{m}_{1,B-2},\hat{m}_{1,B-1}))\right.\nonumber\\
&\ \  \ \ \ \ \ \ \ \ \ \ \ \ \ \ \ \ \ \ \ \ \ \left.\bx_{2}''(\hat{m}''_{2,B-1},\bu(\hat{m}_{1,B-2},\hat{m}'_{1,B-2})),\by''_{B-1}\right\}\in T^{(n)}_\epsilon(UVX_1X_2''Y'').\label{decodPr}
\end{align}
\hrulefill
\vspace*{4pt}
\end{figure*}
Then, in block $B-2$, the decoder has at hand an estimate of the fresh information sent in block $B-2$, namely, $(\hat{m}_{1,B-2},\hat{m}'_{1,B-2})$, and the information sent in block $B-2$ can be decoded next, etc. In general, in block $b$, the decoder has at hand an estimate of the fresh information sent in block $b$, namely, $(\hat{m}_{1,b},\hat{m}'_{1,b})$, and to find the transmitted information in block $b$, the decoder looks for $(\hat{m}_{1,b-1},\hat{m}'_{1,b-1},\hat{m}''_{2,b})$ according to \eqref{decodPrGen}, shown at the top of the next page.
\begin{figure*}[!t]
\normalsize
\setcounter{MYtempeqncnt}{\value{equation}}
\setcounter{equation}{83}
\begin{align}
&\left\{\bu(\hat{m}_{1,b-1},\hat{m}'_{1,b-1}),\bv(\hat{m}_{1,b-1},\hat{m}_{1,b}),\bml{x}_1(\hat{m}'_{1,b},\bu(\hat m_{1,b-1},\hat m'_{1,b-1}),\bv(\hat{m}_{1,b-1},\hat{m}_{1,b}))\right.\nonumber\\
&\ \  \ \ \ \ \ \ \ \ \ \ \ \ \ \ \ \ \ \ \ \ \ \left.\bx_{2}''(\hat{m}''_{2,b},\bu(\hat{m}_{1,b-1},\hat{m}'_{1,b-1})),\by''_{b}\right\}\in T^{(n)}_\epsilon(UVX_1X_2''Y'').\label{decodPrGen}
\end{align}
\hrulefill
\vspace*{4pt}
\end{figure*}

According to the above decoding rule, the decoding of User 1 and User 2 are staggered: at some block $b\in\ppp{1,2,\ldots,B-1}$, the message of User 2 is decoded jointly with the \emph{resolution information} of User 1, and the latter estimates are actually the fresh messages of block $b-1$. 

If in a decoding step (second encoder or the decoder) there is no message index (or no index pair) to satisfy the decoding rule, or if there is more than one index (or index pair), then an index (or an index pair) is chosen at random.

\underline{\emph{Error Analysis:}} The following lemma (see, e.g., \cite[Lemma 4]{PermuterAsnani}) will enable us to bound the probability of error of the super block $nB$ by bounding the probability of error of each block.
\begin{lemma}\label{lem:union}
Let $\ppp{\calA_l}_{l=1}^L$ be a set of events and let $\calA_j^c$ be the complement of the event $\calA_j$. Then,
\begin{align}
\Pr\ppp{\bigcup_{l=1}^L\calA_l}\leq\sum_{l=1}^L\Pr\ppp{\calA_l\vert\calA_1^c,\calA_2^c,\ldots,\calA_{l-1}^c}
\end{align}
where $\calA_{0}=\emptyset$.
\end{lemma}

Using Lemma \ref{lem:union}, we bound the probability of error in the super block $nB$ by the sum of the probability of having an error in each block $b$ given that in previous blocks, the messages were decoded correctly. 

First let us bound the probability that for some $b$, encoder 2 decodes the messages of encoder 1 incorrectly at the end of that block. Using Lemma \ref{lem:union}, it suffices to show that the probability of decoding error in each block goes to zero, assuming that all previous messages in blocks $(1,2,\ldots,b-1)$ were decoded correctly. 

Let $E_{\text{enc},b} = E^{(1)}_{\text{enc},b}\cup E^{(2)}_{\text{enc},b}$ be the event that encoder 2 has an error in decoding $m_{1,b}$ or $m'_{1,b}$. The event $E^{(1)}_{\text{enc},b}$ refers to an error in decoding $m_{1,b}$, while $E^{(2)}_{\text{enc},b}$ refers to an error in decoding $m'_{1,b}$. The term $\Pr\ppp{E_{\text{enc},b}\vert E^c_{\text{enc},b-1}}$ is the probability that encoder 2 incorrectly decoded $m_{1,b}$ or $m'_{1,b}$, given that $m_{1,b-1}$ and $m'_{1,b-1}$ were decoded correctly. We have,
\begin{align}
&\Pr\ppp{E_{\text{enc},b}\vert E^c_{\text{enc},b-1}}\leq \Pr\ppp{E^{(1)}_{\text{enc},b}\vert E^c_{\text{enc},b-1}}\nonumber\\
&\ \ \ \  \ \ \ \ \  \ \ \ \ \ \  \ \ \ \ +\Pr\ppp{E^{(2)}_{\text{enc},b}\vert E^c_{\text{enc},b-1},(E^{(1)}_{\text{enc},b})^c}.
\end{align}
Define the sets
\begin{align}
\calE_{b,m_{1,b}}\triangleq (\bv(m_{1,b-1},m_{1,b}),\bx_{1,b})\in T^{(n)}_\epsilon(VX_1),\label{enc2errev0}
\end{align}
and the set $\calE_{b,m_{1,b}'}$ in \eqref{enc2errev}, shown at the top of the next page, given $m_{1,b-1}$ and $m'_{1,b-1}$. Assume without loss of generality that $m_{1,b-1}=m'_{1,b-1}=m_{1,b}=1$.
\begin{figure*}[!t]
\normalsize
\setcounter{MYtempeqncnt}{\value{equation}}
\setcounter{equation}{87}
\begin{align}
&\calE_{b,m_{1,b}'}\triangleq \left\{\bu(m_{1,b-1},m'_{1,b-1}),\bv(m_{1,b-1},m_{1,b}),\bml{x}_1(m_{1,b}',\bu(m_{1,b-1},m'_{1,b-1}),\bv(m_{1,b-1},m_{1,b})),\bx_{1,b}\right\}\in T^{(n)}_\epsilon(UVX_1X_1).\label{enc2errev}
\end{align}
\hrulefill
\vspace*{4pt}
\end{figure*}
Then, according to \eqref{enc2_1},
\begin{align}
\Pr\ppp{E^{(1)}_{\text{enc},b}\vert E^c_{\text{enc},b-1}} &\leq \Pr\ppp{\bigcup_{m_{1,b}\neq1}\calE_{b,m_{1,b}}\vert E^c_{\text{enc},b-1}}.\label{rhsprob0}
\end{align}
The probability at the right hand side of \eqref{rhsprob0}, is the probability of the event in \eqref{enc2errev0}, given that $m_{1,b-1}$, was decoded correctly. Then, to evaluate \eqref{rhsprob0}, we can equivalently evaluate the probability of the event
\begin{align}
\calE_{b,m_{1,b}}\triangleq (\bv(1,m_{1,b}),\bx_{1,b})\in T^{(n)}_\epsilon(VX_1),
\end{align}
for $m_{1,b}\neq1$. Hence, by classical results, we have,
\begin{align}
\Pr\ppp{E^{(1)}_{\text{enc},b}\vert E^c_{\text{enc},b-1}} &\leq \sum_{m_{1,b}\neq1}\Pr\ppp{\calE_{b,m_{1,b}}\vert E^c_{\text{enc},b-1}}\\
&\leq \sum_{m_{1,b}\neq1}e^{-n(I(V;X_1)-3\epsilon)}\\
&\leq e^{n(R_1-I(V;X_1)+3\epsilon)}.\label{enc2est1}
\end{align}

Next, recall that encoder 2 decodes $m'_{1,b}$ according to \eqref{enc2_2}, given that he already decoded $\hat{m}_{1,b}$ in the first stage, and $\hat{m}_{1,b-1}$ and $\hat{m}'_{1,b-1}$ at the end of block $b-1$. Accordingly, we have,
\begin{align}
&\Pr\ppp{E^{(2)}_{\text{enc},b}\vert E^c_{\text{enc},b-1},(E^{(1)}_{\text{enc},b})^c} \nonumber\\
&= \Pr\ppp{\bigcup_{m_{1,b}'\neq1}\calE_{b,m_{1,b}'}\vert E^c_{\text{enc},b-1},(E^{(1)}_{\text{enc},b})^c}\nonumber\\
&\leq\sum_{m_{1,b}'\neq1}\Pr\ppp{\calE_{b,m_{1,b}'}\vert E^c_{\text{enc},b-1},(E^{(1)}_{\text{enc},b})^c}.\label{rhsprob}
\end{align}
Again, the probability at the right hand side of \eqref{rhsprob}, is the probability of the event in \eqref{enc2errev}, given that $m_{1,b-1}$, $m'_{1,b-1}$, and $m_{1,b}$, were decoded correctly. Then, to evaluate \eqref{rhsprob}, we can equivalently evaluate the probability of the event in \eqref{enc2errevne}, shown at the top of the page, for $m_{1,b}'\neq1$. 
\begin{figure*}[!t]
\normalsize
\setcounter{MYtempeqncnt}{\value{equation}}
\setcounter{equation}{94}
\begin{align}
&\tilde\calE_{b,m_{1,b}'}\triangleq \left\{\bu(1,1),\bv(1,1),\bml{x}_1(m_{1,b}',\bu(1,1),\bv(1,1)),\bx_{1,b}\right\}\in T^{(n)}_\epsilon(UVX_1X_1).\label{enc2errevne}
\end{align}
\hrulefill
\vspace*{4pt}
\end{figure*}
We get
\begin{align}
&\Pr\ppp{\calE_{b,m_{1,b}'}\vert E^c_{\text{enc},b-1},(E^{(1)}_{\text{enc},b})^c}\nonumber\\
&=\sum_{T_{\epsilon}^{(n)}(UVX_1X_1)}P(\bu)P(\bv)P(\bx_1\vert\bu,\bv) P(\bx_1'\vert\bu,\bv)\nonumber\\
&\leq\exp(n(H(U,V,X_1,X_1)+\epsilon))\nonumber\\
&\ \ \ \cdot\exp(-n(H(U,V,X_1)-4\epsilon))\nonumber\\
&\ \ \ \cdot\exp(-n(H(X_1\vert U,V)-\epsilon))\nonumber\\
&= \exp(-n(H(X_1\vert U,V)-6\epsilon)).
\end{align}
Therefore,
\begin{align}
\Pr\ppp{E^{(2)}_{\text{enc},b}\vert E^c_{\text{enc},b-1},(E^{(1)}_{\text{enc},b})^c} &\leq\sum_{m_{1,b}'\neq1}e^{-n(H(X_1\vert U,V)-6\epsilon)}\nonumber\\
&\leq e^{n(R_1'-H(X_1\vert U,V)+6\epsilon)}.\label{enc2est2}
\end{align}
Wrapping up, using \eqref{enc2est1} and \eqref{enc2est2}, by Lemma \ref{lem:union}, if $R_1\leq I(V;X_1)$ and $R_1'\leq H(X_1\vert U,V)$, then encoder 2 can decode all the messages (i.e., over all the $B$ blocks) of encoder 1 correctly, with a probability of error that goes to zero as the block length goes to infinity.

Next, at the receiver side, recall first the decoding rule in \eqref{decodPrGen}, where in block $b$, the decoder looks for $(\hat{m}_{1,b-1},\hat{m}'_{1,b-1},\hat{m}''_{2,b})$ assuming that $(\hat{m}_{1,b},\hat{m}'_{1,b})$ were already decoded in block $b+1$. In the following, we upper bound the overall error probability of the receiver. To this end, we use once again Lemma \ref{lem:union}, as follows. The error probability of the receiver is upper bounded by the sum of the probabilities that in each block $b$ the receiver incorrectly decodes the messages $m_{1,b-1}$, $m'_{1,b-1}$, and $m_{2,b}''$, given that: (1) at block $b+1$ the messages $m_{1,b}$ and $m'_{1,b}$ were decoded correctly, and (2) encoder 2 decoded correctly all the messages of encoder 1 (in all the $B$ blocks). 

Define the event in \eqref{EventErrorDec}, shown at the top of the page, and without loss of generality, assume that $m_{1,b} = m_{1,b}'=1$. Assuming that $m_{1,b-1}=m'_{1,b-1}=m_{2,b}''=1$,
\begin{figure*}[!t]
\normalsize
\setcounter{MYtempeqncnt}{\value{equation}}
\setcounter{equation}{97}
\begin{align}
&E_{m_1,m_1',m_2'',b} \triangleq \left\{\bu(m_1,m_1'),\bv(m_1,{m}_{1,b}),\bml{x}_1({m}'_{1,b},\bu(m_1,m_1'),\bv(m_1,{m}_{1,b})),\right.\left.\bx_{2}''({m}''_2,\bu(m_1,m_1')),\by''_b\right\}\nonumber\\
&\ \ \ \ \ \ \ \ \ \ \ \ \ \ \ \ \ \ \ \ \ \ \ \ \ \ \ \ \ \ \ \ \ \ \ \ \ \ \ \ \ \ \ \ \ \ \ \ \ \ \ \ \ \ \in T^{(n)}_\epsilon(UVX_1X_2''Y'').\label{EventErrorDec}
\end{align}
\hrulefill
\vspace*{4pt}
\end{figure*}
an error occurs if either the correct codewords are not jointly typical with the received sequences, i.e., $E^c_{1,1,1,b}$, or if there exists a different tuple $(m_1,m_1',m_2'')\neq(1,1,1)$ such that $E_{m_1,m_1',m_2'',b}$ occurs. Let $P_{e,b}^{(n)}$ be the decoding error probability at block $b$ given that in blocks $(b+1,\ldots,B)$, there was no decoding error. From the union bound, we obtain that:
\begin{align}
P_{e,b}^{(n)}\leq&\Pr\ppp{E^c_{1,1,1,b}}+\sum_{m_1>1}\Pr\ppp{E_{m_1,1,1,b}}\nonumber\\
&+\sum_{m_1'>1}\Pr\ppp{E_{1,m_1',1,b}}+\sum_{m_2''>1}\Pr\ppp{E_{1,1,m_2'',b}}\nonumber\\
&+\sum_{m_1>1,m_1'>1}\Pr\ppp{E_{m_1,m_1',1,b}}\nonumber\\
&+\sum_{m_1>1,m_2''>1}\Pr\ppp{E_{m_1,1,m_2'',b}}\nonumber\\
&+\sum_{m_1'>1,m_2''>1}\Pr\ppp{E_{1,m_1',m_2'',b}}\nonumber\\
&+\sum_{m_1>1,m_1'>1,m_2''>1}\Pr\ppp{E_{m_1',m_1,m_2'',b}}.\label{decodererror}
\end{align}
Let us upper bound each term in \eqref{decodererror}.
\begin{enumerate}
\item Upper-bounding $\Pr\ppp{E^c_{1,1,1,b}}$: Since we assume that encoder 2 encodes the right messages $m_{1,b-1}$ and $m'_{1,b-1}$ in block $b$, and that the receiver decoded the right messages $m_{1,b}$ and $m'_{1,b}$ at block $b+1$, by the LLN $\Pr\ppp{E^c_{1,1,1,b}}\to0$ as $n\to\infty$.
\item Upper-bounding $\sum_{m_2''>1}\Pr\ppp{E_{1,1,m_2'',b}}$: Let $\calS$ be the set of all sequences $(\bu,\bv,\bx_1,\bx_2'',\by'')$ that belong to $T_{\epsilon}^{(n)}(UVX_1X_2''Y'')$. We then have
\begin{align}
&\Pr\ppp{E_{1,1,m_2'',b}}\nonumber\\
&=\sum_{\calS}P(\bu)P(\bv)P(\bx_1\vert\bu,\bv) P(\bx_2''\vert\bu)\nonumber\\
&\ \ \ \ \ \ \ \ \ \times P(\by''\vert\bu,\bv,\bx_1)\nonumber\\
&\leq\exp(n(H(U,V,X_1,X_2'',Y'')+\epsilon))\nonumber\\
&\ \ \ \cdot\exp(-n(H(U,V,X_1)-4\epsilon))\nonumber\\
&\ \ \ \cdot\exp(-n(H(X_2''\vert U)-\epsilon))\nonumber\\
&\ \ \ \cdot\exp(-n(H(Y''\vert U,V,X_1)-\epsilon))\nonumber\\
& = \exp(-n(I(X_2'';Y''\vert U,V,X_1)-7\epsilon)).
\end{align}
where we have used the fact that $(V,X_1) \markov U \markov X''_2$. Hence, we obtain
\begin{align}
\sum_{m_2''>1}\Pr\ppp{E_{1,1,m_2'',b}}\leq e^{n(R_2''-I(X_2'';Y''\vert U,V,X_1)+7\epsilon)}.\label{upperbounddec}
\end{align}
\item Upper-bounding $\sum_{m_1>1}\Pr\ppp{E_{m_1,1,1,b}}$: We have
\begin{align}
&\Pr\ppp{E_{m_1,1,1,b}}\nonumber\\
&=\sum_{\calS}P(\bu)P(\bv)P(\bx_1\vert\bu,\bv) P(\bx_2''\vert\bu)P(\by'')\nonumber\\
&\leq\exp(n(H(U,V,X_1,X_2'',Y'')+\epsilon))\nonumber\\
&\ \ \ \cdot\exp(-n(H(U,V,X_1)-4\epsilon))\nonumber\\
&\ \ \ \cdot\exp(-n(H(X_2''\vert U)-\epsilon))\nonumber\\
&\ \ \ \cdot\exp(-n(H(Y'')-\epsilon))\nonumber\\
& = \exp(-n(I(U,V,X_1,X_2'';Y'')-7\epsilon)).
\end{align}
where again we use $(V,X_1) \markov U \markov X''_2$. Hence, we obtain
\begin{align}
&\sum_{m_1>1}\Pr\ppp{E_{m_1,1,1,b}}\leq e^{n(R_1-I(U,V,X_1,X_2'';Y'')+7\epsilon)}.\label{upperbounddec2}
\end{align}
\item Upper-bounding $\sum_{m_1'>1}\Pr\ppp{E_{1,m_1',1,b}}$: We have
\begin{align}
&\Pr\ppp{E_{1,m_1',1,b}}\nonumber\\
&=\sum_{\calS}P(\bu)P(\bv)P(\bx_1\vert\bu,\bv) P(\bx_2''\vert\bu)P(\by''|\bv)\nonumber\\
&\leq\exp(n(H(U,V,X_1,X_2'',Y'')+\epsilon))\nonumber\\
&\ \ \ \cdot\exp(-n(H(U,V,X_1)-4\epsilon))\nonumber\\
&\ \ \ \cdot\exp(-n(H(X_2''\vert U)-\epsilon))\nonumber\\
&\ \ \ \cdot\exp(-n(H(Y''|V)-\epsilon))\nonumber\\
& = \exp(-n(I(U,X_1,X_2'';Y''|V)-7\epsilon)).
\end{align}
where again we use $(V,X_1) \markov U \markov X''_2$. Hence, we get
\begin{align}
&\sum_{m_1'>1}\Pr\ppp{E_{1,m_1',1,b}}\leq e^{n(R_1'-I(U,X_1,X_2'';Y''\vert V)+7\epsilon)}.\label{upperbounddec3}
\end{align}
\item Upper-bounding $\sum_{m_1>1,m_1'>1}\Pr\ppp{E_{m_1,m_1',1,b}}$: We have
\begin{align}
&\Pr\ppp{E_{m_1,m_1',1,b}}\nonumber\\
&=\sum_{\calS}P(\bu)P(\bv)P(\bx_1\vert\bu,\bv) P(\bx_2''\vert\bu)P(\by'')\nonumber\\
&\leq\exp(n(H(U,V,X_1,X_2'',Y'')+\epsilon))\nonumber\\
&\ \ \ \cdot\exp(-n(H(U,V,X_1)-4\epsilon))\nonumber\\
&\ \ \ \cdot\exp(-n(H(X_2''\vert U)-\epsilon))\nonumber\\
&\ \ \ \cdot\exp(-n(H(Y'')-\epsilon))\nonumber\\
& = \exp(-n(I(U,V,X_1,X_2'';Y'')-7\epsilon)).
\end{align}
where we use $(V,X_1) \markov U \markov X''_2$. Therefore,
\begin{align}
&\sum_{m_1>1,m_1'>1}\Pr\ppp{E_{m_1,m_1',1,b}}\nonumber\\
&\ \ \ \ \ \ \ \ \ \ \ \leq e^{n(R_1+R_1'-I(U,V,X_1,X_2'';Y'')+7\epsilon)}.\label{upperbounddec4}
\end{align}
\item Upper-bounding $\sum_{m_1>1,m_2''>1}\Pr\ppp{E_{m_1,1,m_2'',b}}$: We have
\begin{align}
&\Pr\ppp{E_{m_1,1,m_2'',b}}\nonumber\\
&=\sum_{\calS}P(\bu)P(\bv)P(\bx_1\vert\bu,\bv) P(\bx_2''\vert\bu)P(\by'')\nonumber\\
&\leq\exp(n(H(U,V,X_1,X_2'',Y'')+\epsilon))\nonumber\\
&\ \ \ \cdot\exp(-n(H(U,V,X_1)-4\epsilon))\nonumber\\
&\ \ \ \cdot\exp(-n(H(X_2''\vert U)-\epsilon))\nonumber\\
&\ \ \ \cdot\exp(-n(H(Y'')-\epsilon))\nonumber\\
& = \exp(-n(I(U,V,X_1,X_2'';Y'')-7\epsilon)).
\end{align}
using $(V,X_1) \markov U \markov X''_2$. Thus,
\begin{align}
&\sum_{m_1>1,m_2''>1}\Pr\ppp{E_{m_1,1,m_2'',b}}\nonumber\\
&\ \ \ \ \ \ \ \ \ \ \ \leq e^{n(R_1+R_2''-I(U,V,X_1,X_2'';Y'')+7\epsilon)}.\label{upperbounddec5}
\end{align}
\item Upper-bounding $\sum_{m_1'>1,m_2''>1}\Pr\ppp{E_{1,m_1',m_2'',b}}$: We have
\begin{align}
&\Pr\ppp{E_{1,m_1',m_2'',b}}\nonumber\\
&=\sum_{\calS}P(\bu)P(\bv)P(\bx_1\vert\bu,\bv) P(\bx_2''\vert\bu)P(\by''|\bv)\nonumber\\
&\leq\exp(n(H(U,V,X_1,X_2'',Y'')+\epsilon))\nonumber\\
&\ \ \ \cdot\exp(-n(H(U,V,X_1)-4\epsilon))\nonumber\\
&\ \ \ \cdot\exp(-n(H(X_2''\vert U)-\epsilon))\nonumber\\
&\ \ \ \cdot\exp(-n(H(Y''|V)-\epsilon))\nonumber\\
& = \exp(-n(I(U,X_1,X_2'';Y''|V)-7\epsilon)).
\end{align}
where the last step follows from $(V,X_1) \markov U \markov X''_2$. Hence, we get
\begin{align}
&\sum_{m_1'>1,m_2''>1}\Pr\ppp{E_{1,m_1',m_2'',b}}\nonumber\\
&\ \ \ \ \ \ \ \ \ \ \ \leq e^{n(R_1'+R_2''-I(U,X_1,X_2'';Y''\vert V)+7\epsilon)}.\label{upperbounddec6}
\end{align}
\item Upper-bounding $\sum_{m_1>1,m_1'>1,m_2''>1}\Pr\ppp{E_{m_1,m_1',m_2'',b}}$: We have
\begin{align}
&\Pr\ppp{E_{m_1,m_1',m_2'',b}}\nonumber\\
&=\sum_{\calS}P(\bu)P(\bv)P(\bx_1\vert\bu,\bv) P(\bx_2''\vert\bu)P(\by'')\nonumber\\
&\leq\exp(n(H(U,V,X_1,X_2'',Y'')+\epsilon))\nonumber\\
&\ \ \ \cdot\exp(-n(H(U,V,X_1)-4\epsilon))\nonumber\\
&\ \ \ \cdot\exp(-n(H(X_2''\vert U)-\epsilon))\nonumber\\
&\ \ \ \cdot\exp(-n(H(Y'')-\epsilon))\nonumber\\
& = \exp(-n(I(U,V,X_1,X_2'';Y'')-7\epsilon)).
\end{align}
where again we use $(V,X_1) \markov U \markov X''_2$. Hence, we obtain
\begin{align}
&\sum_{m_1>1,m_1'>1,m_2''>1}\Pr\ppp{E_{m_1,m_1',m_2'',b}}\nonumber\\
&\ \ \ \ \ \ \ \ \ \ \ \leq e^{n(R_1+R_1'+R_2''-I(U,V,X_1,X_2'';Y'')+7\epsilon)}.\label{upperbounddec7}
\end{align}
\end{enumerate}
Thus, using \eqref{enc2est1}, \eqref{enc2est2}, \eqref{upperbounddec}, \eqref{upperbounddec2}, \eqref{upperbounddec3}, \eqref{upperbounddec4}, \eqref{upperbounddec5}, \eqref{upperbounddec6}, and \eqref{upperbounddec7}, if $(R_1,R_1',R_2'')$ satisfy:
\begin{subequations}\label{allff3}
\begin{IEEEeqnarray}{rCl}
R_1&\leq&I(V;X_1),\label{constt3}\\
R_1'&\leq&H(X_1\vert U,V),\\
R_1&\leq&I(U,V,X_1,X_2'';Y''),\label{constt5}\\
R_1'&\leq&I(U,X_1,X_2'';Y''\vert V),\label{constt6}\\
R_1+R_1'&\leq&I(U,V,X_1,X_2'';Y''),\label{constt7}\\
R_2''&\leq&I(X_2'';Y''\vert U,V,X_1),\\
R_1+R_2''&\leq&I(U,V,X_1,X_2'';Y''),\label{constt1}\\
R_1'+R_2''&\leq&I(U,X_1,X_2'';Y''\vert V),\label{constt4}\\
R_1+R_1'+R_2''&\leq&I(U,V,X_1,X_2'';Y''),\label{constt2}
\end{IEEEeqnarray}
\end{subequations}
then there exists a sequence of codes with a probability of error that goes to zero as the block length goes to infinity. We note to the following simplifications. First, we can remove \eqref{constt5}, \eqref{constt7}, and \eqref{constt1}, due to \eqref{constt2}, and \eqref{constt6} can be removed due to \eqref{constt4}. Second, \eqref{constt4} and \eqref{constt2} can be replaced with $R_1'+R_2''\leq I(X_1,X_2'';Y''\vert V)$ and $R_1+R_1'+R_2''\leq I(X_1,X_2'';Y'')$, respectively, due to the Markov chain $(U,V)\markov(X_1,X_2'')\markov Y''$. Finally, the constraint in \eqref{constt3}, is superfluous due to \eqref{eq:MAC_R1_0}. Indeed,
\begin{align}
I(V;Y\vert X_2)&=H(V\vert X_2)-H(V\vert X_2,Y)\\
&\stackrel{(a)}{\leq} H(V)-H(V\vert X_1,X_2,Y)\\
&\stackrel{(b)}{=}H(V)-H(V\vert X_1)\\
&=I(V;X_1)
\end{align}
where (a) follows from the fact that conditioning reduces entropy, and (b) follows from the Markov chain $(X_2,Y)\markov X_1\markov V$. Thus, to summarize, using the above simplifications, the achievable region for the MAC with unreliable strictly causal cribbing is given (recall \eqref{MAC_R1without})
\begin{subequations}\label{allff3new}
\begin{IEEEeqnarray}{rCl}
R_1&\leq&I(V;Y\vert X_2),\label{constt8}\\
R_2&\leq&I(X_2;Y\vert V),\label{constt9}\\
R_1+R_2&\leq&I(V,X_2;Y),\label{constt10}\\
R_1'&\leq&H(X_1\vert U,V),\label{constt11}\\
R_2''&\leq&I(X_2'';Y''\vert U,V,X_1),\label{constt12}\\
R_1'+R_2''&\leq&I(X_1,X_2'';Y''\vert V),\label{constt13}\\
R_1+R_1'+R_2''&\leq&I(X_1,X_2'';Y''),\label{constt14}
\end{IEEEeqnarray}
\end{subequations}
for some $P_{U,V,X_1,X_2,X_2'',Y,Y''}$ of the form 
\begin{align}
P_UP_{V}P_{X_1|U,V}P_{X_2}P_{X_2''|U}P_{Y|X_1,X_2}P_{Y''|X_1,X_2''},\label{markovChainMAC}
\end{align}
as stated in Theorem \ref{theo:MACstr}.

\subsection{Proof of Theorem \ref{theo:MAC1}}\label{sub:achMac2}

In order to show that all the rate pairs in \eqref{AchCrib} are achievable, we employ Shannon strategies \cite{J_WillemsMeulen85Discrete}. Consider all different strategies (functions), with members $t\in\mathscr{T}\triangleq\calX_2^{\abs{\calX_1}}$ that map inputs $x_1\in\calX_1$ into inputs $x_2''\in\calX_2$. Denote by $t(\cdot)$ the strategy with member $t$ as an operator. 
\begin{definition}
For a DMMAC	$(\calX_1\times\calX_2, P(y''|x_1,x_2''),\calY)$ the DM \emph{derived} MAC is denoted by $(\calX_1\times\mathscr{T}, P^\triangle(y''|x_1,t),\calY)$ where $P^\triangle(y''|x_1,t) \triangleq P(y''|x_1,x_2''=t(x_1))$ for all $x_1\in\calX_1$, $t\in\mathscr{T}$, and $y''\in\calY$. 
\end{definition}

Let $\calR_S$ be the set of rates $(R_1,R_1',R_2,R_2'')$ satisfying
\begin{subequations}
\begin{IEEEeqnarray}{rCl}
R_1&\leq& I(V;Y\vert X_2),\\
R_2&\leq& I(X_2;Y\vert V),\\
R_1+R_2&\leq& I(V,X_2;Y),\\
R_1'&\leq& H(X_1\vert U,V),\\
R_2''&\leq& I(T;Y''\vert U,V,X_1),\\
R_1'+R_2''&\leq& I(X_1,T;Y''\vert V),\\
R_1+R_1'+R_2''&\leq& I(X_1,T;Y''),
\end{IEEEeqnarray}\label{setOf121}
\end{subequations}
for some joint distribution $P(u,v,x_1,x_2,t,y,y'')$ of the form
\begin{align}
&P(u,v,x_1,x_2,t,y,y'') = \nonumber\\ 
&P(u)P(v)P(x_1|u,v)P(x_2)P(t|u)P(y|x_1,x_2)P^\triangle(y''|x_1,t).\label{derivedMAC}
\end{align}
By the achievability scheme for the strictly causal case (Theorem \ref{theo:MACstr}), all rate pairs inside $\calR_S$ are achievable for the above derived MAC. Therefore for the MAC with causal cribbing all rate pairs inside $\calR_S$ must be achievable. If we now restrict the distributions in \eqref{derivedMAC} to satisfy
\begin{align}
P(u,v,x_1,x_2,t,y,y'') &= P(u)P(v)P(x_1|v)P(x_2)P(t)\nonumber\\
&\ \ \ \ \ \cdot P(y|x_1,x_2)P^\triangle(y''|x_1,t),
\end{align}
then
\begin{subequations}
\begin{IEEEeqnarray}{rCl}
H(X_1\vert U,V)&=&H(X_1\vert V),\\
I(T;Y''\vert U,V,X_1)&=&I(X_2'';Y''\vert V,X_1),\\
I(X_1,T;Y''\vert V)&=&I(X_1,X_2'';Y''\vert V),\\
I(X_1,T;Y'')&=&I(X_1,X_2'';Y''),
\end{IEEEeqnarray}\label{allff4}%
\end{subequations}
and\footnote{Recall that for a discrete random variable $X$ with probability mass function $P_X(\cdot)$, the probability mass function $P_Y(\cdot)$ of the discrete random variable $Y=g(X)$ is given by$$
P_Y(y) = \sum_{x:\; y=g(x)}P_X(x).
$$}
\begin{align}
P(v,x_1,x_2'',y'') &= P(v,x_1)\sum_{t:\;t(x_1)=x_2''}P(t)P(y''|x_1,x_2'').
\end{align}
Now, given an arbitrary distribution $P^0(v,x_1,x_2'')=P^0(v,x_1)P^0(x_2''|x_1)$, we note that there always exists a product distribution $P(v,x_1,t) = P(v,x_1)P(t)$ such that
\begin{align}
P(v,x_1)\sum_{t:\;t(x_1)=x_2''}P(t)=P^0(v,x_1,x_2'').
\end{align}
Indeed, this holds for the following choice:
\begin{subequations}
\begin{IEEEeqnarray}{rCl}
P(v,x_1) &=& \sum_{x_2''}P^0(v,x_1,x_2''),\\
P(t) &=& \sum_{x_1'}\frac{P^0(x_1',x_2''=t(x_1'))}{P(x_1')}.
\end{IEEEeqnarray}\label{choose}%
\end{subequations}

Thus, using \eqref{setOf121} and \eqref{allff4}, we conclude that all rate pairs 
\begin{subequations}\label{allff3new1}
\begin{IEEEeqnarray}{rCl}
R_1&\leq&I(V;Y\vert X_2),\label{aall1}\\
R_2&\leq&I(X_2;Y\vert V),\label{aall2}\\
R_1+R_2&\leq&I(V,X_2;Y),\label{aall3}\\
R_1'&\leq&H(X_1\vert V),\label{aall4}\\
R_2''&\leq&I(X_2'';Y''\vert V,X_1),\label{aall5}\\
R_1'+R_2''&\leq&I(X_1,X_2'';Y''\vert V),\label{aall6}\\
R_1+R_1'+R_2''&\leq&I(X_1,X_2'';Y''),\label{aall7}
\end{IEEEeqnarray}
\end{subequations}
for some $P_{V,X_1,X_2,X_2'',Y,Y''}$ of the form 
\begin{align}
P_{V,X_1}P_{X_2}P_{X_2''|X_1}P_{Y|X_1,X_2}P_{Y''|X_1,X_2''},\label{markvochanit43}
\end{align}
are achievable for the MAC with causal cribbing. This completes the proof of Theorem~\ref{theo:MAC1}. 
\qed
\subsection{Proof of Theorem \ref{theo:MAC1out}}\label{sub:achMacout}
We next show that $\calI_{\text{mac}}^{O}$, defined in \eqref{AchCribou}, is an outer bound to the capacity region. We start with a sequence of codes $(n,e^{nR_1},e^{nR_1'},e^{nR_2},e^{nR_2''},\epsilon_n)$ with increasing blocklength $n$, satisfying $\lim_{n\to\infty}\epsilon_n=0$. We denote by $M_k$ the random message from $\calN_k$, for $k=1,2$, and by $M_1'$ and $M_2''$ the messages from $\calN_1'$ and $\calN_2''$, respectively. If the cribbing is absent, by Fano's inequality we can bound the rate $R_1$ as follows
\begin{align}
nR_1&-n\delta_n\leq I\p{M_1;Y^n\vert M_2}\\
& = \sum_{i=1}^nI\p{M_1;Y_i\vert Y^{i-1},M_2}\\
& \stackrel{(a)}{\leq} \sum_{i=1}^nI\p{M_1,Y^{i-1};Y_i\vert M_2}\\
& \stackrel{(b)}{=} \sum_{i=1}^nI\p{M_1;Y_i\vert M_2} + I(Y^{i-1};Y_i\vert M_1,M_2)\\
& \stackrel{(c)}{=} \sum_{i=1}^nI\p{M_1;Y_i\vert M_2,X_{2,i}} + I(Y^{i-1};Y_i\vert M_1,M_2)\\
& \stackrel{(a)}{\leq}   \sum_{i=1}^nI\p{M_1,M_2;Y_i\vert X_{2,i}} + I(Y^{i-1};Y_i\vert M_1,M_2)\\
& \stackrel{(b)}{=} \sum_{i=1}^nI\p{M_1;Y_i\vert X_{2,i}}+I\p{M_2;Y_i\vert M_1,X_{2,i}}\nonumber\\
&\hspace{1.5cm}+ I(Y^{i-1};Y_i\vert M_1,M_2)\\
& \stackrel{(d)}{=} \sum_{i=1}^n I\p{M_1;Y_i\vert X_{2,i}} + I(Y^{i-1};Y_i\vert M_1,M_2)
\label{1qelou}
\end{align}
where $\lim_{n\to\infty}\delta_n=0$, due to $\lim_{n\to\infty}\epsilon_n=0$, (a) follows from the chain rule for mutual information and the non-negativity of the mutual information, (b) follows from the chain rule for mutual information, (c) is due to the fact that $X_{2,i}$ is a deterministic function of $M_2$, and (d) follows from the Markov chain $M_2\markov\p{M_1,X_{2,i}}\markov Y_i$, proved in Appendix \ref{app:1} (see, Lemma \ref{ref:auxMark}). Thus, $I\p{M_2;Y_i\vert M_1,X_{2,i}}=0$. Continuing, note that $I(Y^{i-1};Y_i\vert M_1,M_2)$, appearing in \eqref{1qelou}, can be upper bounded as follows 
\begin{align}
&I(Y^{i-1};Y_i\vert M_1,M_2) \stackrel{(a)}{=} I(Y^{i-1};Y_i\vert M_1,M_2,X_2^i)\\
&\hspace{2cm}\leq I(Y^{i-1},X_1^{i-1};Y_i\vert M_1,M_2,X_2^i)\\
&\hspace{2cm}\stackrel{(b)}{=} I(X_1^{i-1};Y_i\vert M_1,M_2,X_2^i)\\
&\hspace{2cm}\leq I(X_2^{i-1},M_2,X_1^{i-1};Y_i\vert M_1,X_{2i})\\
&\hspace{2cm}\stackrel{(c)}{=} I(X_1^{i-1};Y_i\vert M_1,X_{2,i})\nonumber\\
&\hspace{2.5cm}+I(X_{2}^{i-1},M_2;Y_i\vert M_1,X_{2,i},X_1^{i-1})\\
&\hspace{2cm}\stackrel{(d)}{=} I(X_1^{i-1};Y_i\vert M_1,X_{2,i})\label{recallr}
\end{align}
where (a) is due to the fact that $X_2^i$ is a deterministic function of $M_2$, (b) follows from the fact that $Y^{i-1}\markov(X_1^{i-1},X_2^{i},M_1,M_2)\markov Y_i$ (see, Lemma \ref{ref:auxMark}), (c) follows from the chain rule of mutual information, and finally (d) is due to the Markov chain $(M_2,X_2^{i-1})\markov\p{M_1,X_1^{i-1},X_{2,i}}\markov Y_i$ (see, Lemma \ref{ref:auxMark}). Wrapping up, we obtained
\begin{align}
nR_1-n\delta_n&\leq\sum_{i=1}^n I\p{M_1;Y_i\vert X_{2,i}} + I(X_1^{i-1};Y_i\vert M_1,X_{2,i})\nonumber\\
& = \sum_{i=1}^n I\p{M_1,X_1^{i-1};Y_i\vert X_{2,i}}.\label{end1}
\end{align}
Next, for $R_2$ we have:
\begin{align}
nR_2&-n\delta_n\leq I(M_2;Y^n\vert M_1,M_1')\\
&=\sum_{i=1}^nI(M_2;Y_i\vert M_1,M_1',Y^{i-1})\\
& \stackrel{(a)}{=}\sum_{i=1}^nI(X_{2,i},M_2;Y_i\vert M_1,M_1',X_{1,i},Y^{i-1})\\
& \leq \sum_{i=1}^nI(X_{2,i},M_1,M_1',M_2,Y^{i-1};Y_i\vert X_{1,i})\\
&  \stackrel{(b)}{=} \sum_{i=1}^nI(X_{2,i};Y_i\vert X_{1,i})\nonumber\\
&\hspace{1.5cm}+ I(M_1,M_1',M_2,Y^{i-1};Y_i\vert X_{1,i},X_{2,i})\\
& \stackrel{(c)}{=} \sum_{i=1}^nI(X_{2,i};Y_i\vert X_{1,i})\label{end2}
\end{align}
where (a) follows from the fact that $X_{2,i}$ and $X_{1,i}$ are deterministic functions of $M_2$ and $(M_1,M_1')$, respectively, (b) is due to the chain rule for mutual information, and (c) follows from the Markov chain $(M_1,M_1',M_2,Y^{i-1})\markov(X_{1,i},X_{2,i})\markov Y_i$. Finally, for the sum rate we have
\begin{align}
n\p{R_1+R_2}&-n\delta_n\leq \sum_{i=1}^nI\p{M_1,M_2;Y_i\vert Y^{i-1}}\\
&\leq \sum_{i=1}^nI\p{M_1,M_2,Y^{i-1};Y_i}\\
& = \sum_{i=1}^nI\p{M_1,M_2;Y_i}+I(Y^{i-1};Y_i\vert M_1,M_2)
\end{align}
where the last equality follows from the chain rule. However, we already saw that (recall \eqref{recallr}):
\begin{align}
I(Y^{i-1};Y_i\vert M_1,M_2)\leq I(X_1^{i-1};Y_i\vert M_1,X_{2,i}),
\end{align}
and thus
\begin{align}
n\p{R_1+R_2}-n\delta_n&\leq \sum_{i=1}^nI\p{M_1,M_2;Y_i}\nonumber\\
&\ \ \ \ \ \ \ +I(X_1^{i-1};Y_i\vert M_1,X_{2,i})\\
& \stackrel{(a)}{=} \sum_{i=1}^nI\p{M_1,M_2,X_{2,i};Y_i}\nonumber\\
&\ \ \ \ \ \ \ +I(X_1^{i-1};Y_i\vert M_1,X_{2,i})\\
& \stackrel{(b)}{=} \sum_{i=1}^nI\p{M_1,X_{2,i};Y_i}\nonumber\\
&\ \ \ \ \ \ \ +I(X_1^{i-1};Y_i\vert M_1,X_{2,i})\\
& = \sum_{i=1}^nI\p{M_1,X_1^{i-1},X_{2,i};Y_i}\label{end3}
\end{align} 
where in (a) we use the fact that $X_{2,i}$ is a deterministic function of $M_2$, and (b) is due to the fact that $I\p{M_1,M_2,X_{2,i};Y_i} = I\p{M_1,X_{2,i};Y_i}+I\p{M_2;Y_i|M_1,X_{2,i}}$ and that $M_2\markov (M_1,X_{2,i})\markov Y_i$. 

Now, when cribbing is present, by Fano's inequality we bound the rate $R_1'$ as follows:
\begin{align}
nR_1'&-n\delta_n\leq I(M_1';Y^{n''}\vert M_1)\\
& = I(M_1,M_1';Y^{n''}\vert M_1)\\
& \stackrel{(a)}{=} I(M_1,M_1',X_1^n;Y^{n''}\vert M_1)\\
& \stackrel{(b)}{=} I(X_1^n;Y^{n''}\vert M_1)+ I(M_1,M_1';Y^{n''}\vert M_1,X_1^n)\\
& \stackrel{(c)}{=} I(X_1^n;Y^{n''}\vert M_1)\\
& \leq H(X_1^n\vert M_1)\\
&\stackrel{(d)}{=}\sum_{i=1}^nH(X_{1,i}\vert M_1,X_1^{i-1})
\label{end4}
\end{align}
where (a) follows the fact that $X_1^n$ is a deterministic function of $(M_1,M_1')$, (b) is due to the chain rule for mutual information, (c) follows from the Markov chain $(M_1,M_1')\markov X_1^n\markov Y^{n''}$ (see, Lemma \ref{ref:auxMark}), and (d) is due to the entropy chain rule. Next, for $R_2''$ we have:
\begin{align}
nR_2''&-n\delta_n\leq I(M_2'';Y^{n''}\vert M_1,M_1')\\
& = \sum_{i=1}^nI(M_2'';Y_i''\vert Y^{i-1''},M_1,M_1')\\
& \stackrel{(a)}{=} \sum_{i=1}^nI(M_2'';Y_i''\vert Y^{i-1''},M_1,M_1',X_1^{i-1},X_{1,i})\\
& \leq \sum_{i=1}^nI(Y^{i-1''},M_1',M_2'';Y_i''\vert M_1,X_1^{i-1},X_{1,i})\\
& \stackrel{(b)}{=} \sum_{i=1}^nI(Y^{i-1''},M_1',M_2'',X_{2,i}'';Y_i''\vert M_1,X_1^{i-1},X_{1,i})\\
& \stackrel{(c)}{=} \sum_{i=1}^nI(X_{2,i}'';Y_i''\vert  M_1,X_1^{i-1},X_{1,i})
\label{end5}
\end{align}
where (a) is due to the fact that $X_1^{i}$ is a deterministic function of $M_1$ and $M_1'$, (b) follows the fact that $X_{2,i}''$ is a deterministic function of $(M_2'',X_1^i)$, and (c) follows from the chain rule for mutual information and the Markov chain $(M_1,X_1^{i-1},Y^{i-1''},M_1',M_2'')\markov(X_{1,i},X_{2,i}'') \markov Y_i''$. Finally, for the sum rate $R_1+R_1'+R_2''$, we have:
\begin{align}
n(R_1+R_1'+R_2'')-n\delta_n&\leq I(M_1,M_1',M_2'';Y^{n''})\\
&\leq \sum_{i=1}^nI(X_{1,i},X''_{2,i};Y_i'').\label{end7}
\end{align}
So, hitherto we have that:
\begin{subequations}
\begin{align}
        &n(R_1-\delta_n)\leq \sum_{i=1}^n I(M_1,X_1^{i-1};Y_i\vert X_{2,i})\label{ee1elou}\\
				&n(R_2-\delta_n)\leq \sum_{i=1}^n I(X_{2,i};Y_i\vert X_{1,i})\label{ee2elou}\\
				&n(R_1+R_2-\delta_n)\leq \sum_{i=1}^n I(M_1,X_1^{i-1},X_{2,i};Y_i)\label{ee3elou}\\
				&n(R_1'-\delta_n)\leq \sum_{i=1}^n H(X_{1i}\vert M_1,X_1^{i-1})\label{ee4elou}\\
				&n(R_2''-\delta_n)\leq \sum_{i=1}^n I(X_{2,i}'';Y_i''\vert M_1,X_1^{i-1},X_{1,i})\label{ee5elou}\\
				&n(R_1'+R_2''-\delta_n)\leq \sum_{i=1}^n H(X_{1i}\vert M_1,X_1^{i-1})\nonumber\\
				&\ \ \ \ \ \ \ \ \ \ \ \ \ \ \ \ \ \ \ \ \ \ \ +I(X_{2,i}'';Y_i''\vert M_1,X_1^{i-1},X_{1,i})\label{ee6elou}\\
				&n(R_1+R_1'+R_2''-\delta_n)\leq \sum_{i=1}^nI(X_{1,i},X_{2,i}'';Y_i'')\label{ee7elou}.
\end{align}
\end{subequations}
We are now in a position to define our auxiliary RV. From \eqref{ee1elou}-\eqref{ee7elou}, letting $V_i\triangleq\p{M_1,X_1^{i-1}}$, and thus preserving the Markov chain induced by $\calP$, we have that
\begin{subequations}\label{finalelou}
\begin{align}
        &n(R_1-\delta_n)\leq \sum_{i=1}^n I(V_i;Y_i\vert X_{2,i})\\
				&n(R_2-\delta_n)\leq \sum_{i=1}^n I(X_{2,i};Y_i\vert X_{1,i})\\
				&n(R_1+R_2-\delta_n)\leq \sum_{i=1}^n I(V_i,X_{2,i};Y_i)\\
				&n(R_1'-\delta_n)\leq \sum_{i=1}^n H(X_{1i}\vert V_i)\\
				&n(R_2''-\delta_n)\leq \sum_{i=1}^n I(X_{2,i}'';Y_i''\vert V_i,X_{1,i})\\
				&n(R_1'+R_2''-\delta_n)\leq \sum_{i=1}^n H(X_{1i}\vert V_i)\nonumber\\
				&\ \ \ \ \ \ \ \ \ \ \ \ \ \ \ \ \ \ \ \ \ \ \ \ \ \ \ \ +I(X_{2,i}'';Y_i''\vert V_i,X_{1,i})\\
				&n(R_1+R_1'+R_2''-\delta_n)\leq \sum_{i=1}^nI(X_{1,i},X_{2,i}'';Y_i'').
\end{align}
\end{subequations}
Using the standard time-sharing argument as in \cite[Ch. 14.3]{CoverThomas}, one can rewrite \eqref{finalelou} by introducing an appropriate time-sharing random variable. Therefore, if $\epsilon_n\to0$ as $n\to\infty$, the convex hull of this region can be shown to be equivalent to the convex hull of the region in \eqref{AchCribou}.

\begin{remark}
As was mentioned in the paragraph preceding Theorem \ref{theo:MAC1out}, one can obtain the same outer bound also for the case of non-causal cribbing (see, \eqref{eq:MAC_enc_3strnon}). Indeed, it is evident that the only places where the casual assumption play a role are in the bounds on $R_2''$ and $R_1+R_1'+R_2''$. It is easy to see that the bound on $R_1+R_1'+R_2''$ will not change, and regarding $R_2''$, we have (see, \eqref{end5}):
\begin{align}
nR_2''&-n\delta_n\leq I(M_2'';Y^{n''}\vert M_1,M_1')\\
& = \sum_{i=1}^nI(M_2'';Y_i''\vert Y^{i-1''},M_1,M_1')\\
& \stackrel{(a)}{=} \sum_{i=1}^nI(M_2'';Y_i''\vert Y^{i-1''},M_1,M_1',X_1^{n})\\
& \leq \sum_{i=1}^nI(M_1,X_1^{n/i},Y^{i-1''},M_1',M_2'';Y_i''\vert X_{1,i})\\
& \stackrel{(b)}{=} \sum_{i=1}^nI(M_1,X_1^{n/i},Y^{i-1''},M_1',M_2'',X_{2,i}'';Y_i''\vert X_{1,i})\\
& \stackrel{(c)}{=} \sum_{i=1}^nI(X_{2,i}'';Y_i''\vert  X_{1,i})
\label{end5non}
\end{align}
where (a) is due to the fact that $X_1^{n}$ is a deterministic function of $M_1$ and $M_1'$, (b) follows the fact that $X_{2,i}''$ is a deterministic function of $(M_2'',X_1^n)$, and (c) follows from the Markov chain $(M_1,X_1^{n/i},Y^{i-1''},M_1',M_2'')\markov(X_{1,i},X_{2,i}'') \markov Y_i''$, where $X^{n/i} = (X^{i-1},X_{i+1}^n)$.
\end{remark}
\qed

\appendices
\numberwithin{equation}{section}
\section{Auxiliary Markov Chains Relations}
\label{app:1}
\begin{lemma}\label{ref:auxMark}
The following relations hold:
\begin{enumerate}
\item $M_2\markov\p{M_1,X_{2,i}}\markov Y_i$
\item $(M_2,X_2^{i-1})\markov\p{M_1,X_1^{i-1},X_{2,i}}\markov Y_i$
\item $Y^{i-1}\markov\p{X_1^{i-1},X_2^{i-1}}\markov Y_i$
\item $Y^{i-1}\markov\p{X_1^{i-1},X_2^{i-1},M_1,M_2}\markov Y_i$
\item $Y^{i-1}\markov\p{X_1^{i-1},X_2^{i},M_1,M_2}\markov Y_i$
\item $(M_1,M_1')\markov X_1^n\markov Y^{n''}$
\end{enumerate}
\end{lemma}

\emph{Proof of Lemma \ref{ref:auxMark}}: 
First, recall that:
\begin{align}
\p{M_1,M_2,Y^{i-1},X_1^{i-1},X_2^{i-1}}\markov\p{X_{1,i},X_{2,i}}\markov Y_i.\label{MarkovChel}
\end{align}
Thus, the first item of Lemma \ref{ref:auxMark} follows from:
\begin{align}
P_{Y_i\vert M_1,X_{2,i},M_2} &= \sum_{x_{1,i}} P_{Y_i\vert M_1,X_{2,i},M_2,X_{1,i}}\nonumber\\
&\ \ \ \ \ \ \ \ \ \ \ \times P_{X_{1i}\vert M_1,X_{2,i},M_2}\label{Mar1el}\\
&= \sum_{x_{1,i}} P_{Y_i\vert M_1,X_{2,i},X_{1,i}}P_{X_{1i}\vert M_1,X_{2,i}}\label{Mar2el}\\
& = P_{Y_i\vert M_1,X_{2,i}},\label{Mar3el}
\end{align}
where in the second equality we have used \eqref{MarkovChel}, and the fact that $X_1$ is independent of $M_2$. The second item of Lemma \ref{ref:auxMark} follows exactly in the same way as above. Indeed,
\begin{align}
P_{Y_i\vert M_1,X_{2}^i,M_2,X_1^{i-1}} &= \sum_{x_{1,i}} P_{Y_i\vert M_1,X_{2}^i,M_2,X_1^{i}}\nonumber\\
&\ \ \ \ \times P_{X_{1i}\vert M_1,X_{2}^i,M_2,X_1^{i-1}}\\
&= \sum_{x_{1,i}} P_{Y_i\vert M_1,X_1^i,X_{2,i}}\nonumber\\
&\ \ \ \ \ \ \times P_{X_{1i}\vert M_1,X_1^{i-1},X_{2,i}}\\
& = P_{Y_i\vert M_1,X_1^{i-1},X_{2,i}}.
\end{align}
Next, the third item is true because:
\begin{align}
P_{Y_i\vert X_1^{i-1},X_2^{i-1},Y^{i-1}} &= \sum_{x_{1,i},x_{2,i}} P_{Y_i\vert X_1^{i-1},X_2^{i-1},Y^{i-1},X_{1,i},X_{2,i}}\nonumber\\
&\ \ \ \ \ \ \ \ \times P_{X_{1i},X_{2i}\vert X_1^{i-1},X_2^{i-1},Y^{i-1}}\\
&= \sum_{x_{1,i},x_{2,i}} P_{Y_i\vert X_1^{i},X_2^{i}}P_{X_{1i},X_{2i}\vert X_1^{i-1},X_2^{i-1}}\\
&= \sum_{x_{1,i},x_{2,i}} P_{X_{1,i},X_{2,i},Y_i\vert X_1^{i-1},X_2^{i-1}}\\
& = P_{Y_i\vert X_1^{i-1},X_2^{i-1}}
\end{align}
where the second equality follows from the fact that the channel is memoryless and the fact that there is no feedback. The forth item follows in exactly the same way. The fifth item follows from:
\begin{align}
&P_{Y_i\vert X_1^{i-1},X_2^{i},Y^{i-1},M_1,M_2} \nonumber\\
&= \sum_{x_{1,i}} P_{Y_i\vert X_1^{i-1},X_2^{i},Y^{i-1},X_{1,i},M_1,M_2}\nonumber\\
&\ \ \ \ \ \ \ \ \times P_{X_{1i}\vert X_1^{i-1},X_2^{i},Y^{i-1},M_1,M_2}\\
&= \sum_{x_{1,i}} P_{Y_i\vert X_1^{i},X_2^{i},M_1,M_2}P_{X_{1i}\vert X_1^{i-1},X_2^{i},M_1,M_2}\\
&= \sum_{x_{1,i}} P_{X_{1,i},Y_i\vert X_1^{i-1},X_2^{i},M_1,M_2}\\
& = P_{Y_i\vert X_1^{i-1},X_2^{i},M_1,M_2}
\end{align}
where again the second equality follows from the fact that the channel is memoryless and the fact that there is no feedback. Finally, we obtain the sixth item due to the same reasons:
\begin{align}
&P_{Y^{n''}\vert X_1^n,M_1,M_1'} \nonumber\\
&= \sum_{x_2^{n''}} P_{Y^{n''}\vert X_1^n,X_2^{n''},M_1,M_1'}P_{X_2^{n''}\vert X_1^n,M_1,M_1'}\\
&= \sum_{x_2^{n''}} P_{Y^{n''}\vert X_1^n,X_2^{n''}}P_{X_2^{n''}\vert X_1^n}\\
&= \sum_{x_{1,i}} P_{Y^{n''},X_2^{n''}\vert X_1^n}\\
& = P_{Y^{n''}\vert X_1^n}.
\end{align}
 \qed

\section{Proof of Lemma \ref{lem1p}}
\label{app:2}
\emph{Proof:} In the following, we upper bound each constraint in \eqref{AchCrib2}, and show that that the upper bounds can be achieved by taking $V=X_1$. We have:
\begin{align}
R_1 &\leq I(V;Y|X_2)\\
&\leq I(V,X_1;Y|X_2)\\
& = I(X_1;Y\vert X_2),\label{upm1}
\end{align}
where we have used the fact that $V\markov (X_1,X_2)\markov Y$. Next,
\begin{align}
R_2 &\leq I(X_2;Y|V)\\
& = H(X_2\vert V) - H(X_2\vert V,Y)\\
& \leq H(X_2\vert X_1) -H(X_2\vert X_1,Y)\\
& = I(X_2;Y|X_1)\label{upm2}
\end{align}
where the inequality follows from the fact that $X_2$ is independent of $(V,X_1)$, and the fact that:
\begin{align}
H(X_2\vert X_1,Y) &= H(X_2\vert X_1,V,Y)\\
&\leq H(X_2\vert V,Y)
\end{align}
where the inequality is due to the fact that conditioning reduces entropy, and the equality follows from the relation $V\markov(X_1,Y)\markov X_2$. Indeed, first note that:
\begin{align}
P_{X_2,V\vert X_1,Y} &= \frac{P_{X_1X_2Y}P_{V\vert X_1,X_2,Y}}{P_{X_1,Y}}\\
& = P_{X_2\vert X_1,Y}P_{V\vert X_1,X_2,Y}\\
& = P_{X_2\vert X_1,Y}P_{V\vert X_1,X_2}\\
& = P_{X_2\vert X_1,Y}P_{V\vert X_1} \\
&= P_{X_2\vert X_1,Y}P_{V\vert X_1,Y}
\end{align}
where the third and last equalities follow from the relations $V\markov(X_1,X_2)\markov Y$ and $V\markov X_1\markov Y$, respectively, which are true due to \eqref{eq:MAC_joint}. For the sum rate, we have:
\begin{align}
R_1+R_2&\leq I(V,X_2;Y)\\
&\leq I(V,X_1,X_2;Y)\\
& = I(X_1,X_2;Y)\label{upm3}
\end{align}
in which the last equality follow from $V\markov(X_1,X_2)\markov Y$. Similarly, for $R_2''$, we obtain:
\begin{align}
R_2''&\leq I(X_2'';Y''\vert X_1,V)\\
& = H(Y''\vert X_1,V) - H(Y''\vert X_1,V,X_2'')\\
&\leq H(Y''\vert X_1) - H(Y''\vert X_1,X_2'')\\
& = I(X_2'';Y''\vert X_1)\label{upm4}
\end{align}
where the inequality follows from the fact that conditioning reduces entropy, and the relation $V\markov(X_1,X_2'')\markov Y''$. Finally, the result follows by noticing that the obtained upper bounds in \eqref{upm1}, \eqref{upm2}, \eqref{upm3}, and \eqref{upm4} are independent of $V$, and can be achieved by taking $V=X$.
\qed

\section{The Capacity Region in Example \ref{exmp:ex3}}
\label{app:3}

First, note that for $i\in\ppp{0,1}$ and $j\in\ppp{0,1,2,3}$:
\begin{align}
P_{Y_1}(0) &= P_{X_1}(1-q)+\bar{P}_{X_1}q,\\
\nu_i&\triangleq\Pr\ppp{X_1=0|Y_1=i}\nonumber\\
 \ \ \ &= \frac{P_{X_1}(1-q)^{1-i}q^i}{P_{X_1}(1-q)^{1-i}q^i+\bar{P}_{X_1}(1-q)^{i}q^{1-i}},\\
\lambda_{ij}&\triangleq\Pr\ppp{Y_2=0|Y_1=i,X_2=j}\nonumber\\
&=\nu_i\pp{\delta\p{j}+\frac{1}{2}\pp{\delta\p{j-2}+\delta\p{j-3}}}\nonumber\\
& \ \ \ + \bar{\nu}_i\pp{\delta\p{j-2}+\frac{1}{2}\pp{\delta\p{j}+\delta\p{j-1}}},
\end{align}
and
\begin{align}
\Pr\ppp{Y_2=0|Y_1=i} = \sum_{j=0}^3p_j\lambda_{ij}.
\end{align}
Then, it is easy to check that:
\begin{align}
H(Y_2|X_2,Y_1) &= \sum_{i,j}P_{Y_1}(i)p_j\calH_2(\lambda_{ij}),\label{entropy1}\\
H(Y_2|Y_1) &= \sum_{i=0}^1P_{Y_1}(i)\calH_2\p{\sum_j\lambda_{ij}},\\
H(Y_2|X_1,Y_1) &= H(Y_2|X_1)\nonumber\\
& = P_{X_1}\calH_2\p{p_0+\frac{1}{2}\p{p_2+p_3}}\nonumber\\
&\ \ \ \ +\bar{P}_{X_1}\calH_2\p{p_2+\frac{1}{2}\p{p_0+p_1}}\label{entropy2}.
\end{align}
Using the above results and \eqref{AchCrib3}, we have have:
\begin{align}
R_1&\leq I(X_1;Y_1,Y_2|X_2)\nonumber\\
& = H(Y_1,Y_2|X_2)-H(Y_1,Y_2|X_1,X_2)\nonumber\\
& = H(Y_2)+H(Y_2|X_2,Y_1)-H(Y_1|X_1)-H(Y_2|X_1,X_2)\nonumber\\
& = \calH_2(P_{X_1}\ast q)-\calH_2(q) + H(Y_2|X_2,Y_1)\nonumber\\
& \ \ \ \ \ -P_{X_1}(P_2+P_3)-\bar{P}_{X_1}(P_0+P_1)\nonumber\\
&\triangleq \calR_1,
\end{align}
and
\begin{align}
R_2\leq H(Y_2|X_1,Y_1)-P_{X_1}(P_2+P_3)-\bar{P}_{X_1}(P_0+P_1).
\end{align}
For the sum rate, we get:
\begin{align}
R_1+R_2&\leq I(X_1,X_2;Y_1,Y_2)\nonumber\\
& = I(X_1;Y_1,Y_2|X_2) + I(X_2;Y_1,Y_2)\nonumber\\
& = \calR_1+H(Y_2|Y_1)-H(Y_2|X_2,Y_1).
\end{align}
Regarding $R_2''$, choosing the distribution $P_{X_2''|X_1}$ as in \eqref{X2primedis}-\eqref{X2primedis2}, we readily get that
\begin{align}
R_2''\leq 1,
\end{align}
and
\begin{align}
R_1+R_2''&\leq I(X_1,X_2'';Y_1,Y_2)\nonumber\\
& = I(X_1;Y_1,Y_2)+I(X_2'';Y_1,Y_2|X_1)\nonumber\\
& = I(X_1;Y_1)+I(X_1;Y_2|Y_1)+1\nonumber\\
& = 1+\calH_2(P_{X_1}\ast q)-\calH_2(q).
\end{align}
Therefore, we have obtain that the capacity region in Example \ref{exmp:ex3} is:
\begin{subequations}\label{exmp3cap}
\begin{IEEEeqnarray}{rCl}
R_1 &\leq& \calR_1,\\
R_2 &\leq& H(Y_2|X_1,Y_1)-P_{X_1}(P_2+P_3)\nonumber\\
&&-\bar{P}_{X_1}(P_0+P_1),\\
R_1+R_2 &\leq& \calR_1+H(Y_2|Y_1)-H(Y_2|X_2,Y_1),\\
R_2''&\leq&1,\\
R_1+R_2''&\leq&1+\calH_2(P_{X_1}\ast q)-\calH_2(q),
\end{IEEEeqnarray}%
\end{subequations}
where $H(Y_2|X_1,Y_1)$, $H(Y_2|Y_1)$, and $H(Y_2|X_2,Y_1)$, are given in \eqref{entropy1}-\eqref{entropy2}.


\end{document}

%% file: myshorts.tex
\newcommand {\bu} {\mbox{\boldmath $u$}}
\newcommand {\bv} {\mbox{\boldmath $v$}}

\newcommand {\bx} {\mbox{\boldmath $x$}}
\newcommand {\by} {\mbox{\boldmath $y$}}

\newcommand {\bE} {\mathbb{E}}

\newcommand{\calA}{{\cal A}}

\newcommand{\calE}{{\cal E}}

\newcommand{\calH}{{\cal H}}
\newcommand{\calI}{{\cal I}}

\newcommand{\calN}{{\cal N}}

\newcommand{\calP}{{\cal P}}

\newcommand{\calR}{{\cal R}}
\newcommand{\calS}{{\cal S}}

\newcommand{\calU}{{\cal U}}
\newcommand{\calV}{{\cal V}}

\newcommand{\calX}{{\cal X}}
\newcommand{\calY}{{\cal Y}}

\newcommand{\be}{\begin{equation}}
\newcommand{\ee}{\end{equation}}
\newcommand{\beqna}{\begin{eqnarray}}
\newcommand{\eeqna}{\end{eqnarray}}
